\documentclass[fleqn,usenatbib,useAMS]{mnras}
\usepackage[T1]{fontenc}
\usepackage{ae,aecompl}
\usepackage{footnote}

\usepackage{graphicx}	
\usepackage{amsmath}	
\usepackage{amssymb}	
\usepackage{multicol}        
\usepackage{bm}		
\usepackage{pdflscape}	

\DeclareRobustCommand{\VAN}[3]{#2}
\let\VANthebibliography\thebibliography
\def\thebibliography{\DeclareRobustCommand{\VAN}[3]{##3}\VANthebibliography}

\usepackage{graphicx}	
\usepackage{amsmath,bm}	
\usepackage{amssymb}	
\usepackage{lmodern}
\usepackage{array, makecell}
\usepackage[utf8]{inputenc}

\usepackage{tikz,xcolor,hyperref}

\definecolor{lime}{HTML}{A6CE39}
\DeclareRobustCommand{\orcidicon}{
	\begin{tikzpicture}
	\draw[lime, fill=lime] (0,0) 
	circle [radius=0.16] 
	node[white] {{\fontfamily{qag}\selectfont \tiny ID}};
	\draw[white, fill=white] (-0.0625,0.095) 
	circle [radius=0.007];
	\end{tikzpicture}
	\hspace{-2mm}
}

\foreach \x in {A, ..., Z}{\expandafter\xdef\csname orcid\x\endcsname{\noexpand\href{https://orcid.org/\csname orcidauthor\x\endcsname}
			{\noexpand\orcidicon}}
}

\def\mnras{MNRAS}
\def\apj{ApJ}
\def\apjs{ApJS}
\def\aap{A\&A}
\def\aaps{A\&AS}
\def\aj{AJ}
\def\pasp{PASP}

\newcommand{\ha}{H$\alpha$}

\newcommand{\oiii}{[O\,{\sc iii}]}

\newcommand{\nii}{[N\,{\sc ii}]}

\newcommand{\niiab}{[N\,{\sc ii}]\ 6548,\ 6584\,\AA}

\def \st{\ifmmode{^{\mathrm{st}}}\else{${^{\mathrm{st}}}$}\fi}
\def \nd{\ifmmode{^{\mathrm{nd}}}\else{${^{\mathrm{nd}}}$}\fi}
\def \rd{\ifmmode{^{\mathrm{rd}}}\else{${^{\mathrm{rd}}}$}\fi}
\def \th{\ifmmode{^{\mathrm{th}}}\else{${^{\mathrm{th}}}$}\fi}
\usepackage[normalem]{ulem}  
\newcommand\redout{\bgroup\markoverwith
{\textcolor{red}{\rule[0.5ex]{2pt}{0.8pt}}}\ULon}

\newcommand{\hnii}{{\rm H}$\alpha$+[N~{\sc ii}]}
\newcommand{\h}{$^{\rm h}$}
\newcommand{\m}{$^{\rm m}$}
\newcommand{\s}{$^{\rm s}$}

\newcommand{\flux}{$10^{-12}$ erg s$^{-1}$ cm$^{-2}$}
\newcommand{\fluxb}{$10^{-13}$ erg s$^{-1}$ cm$^{-2}$}
\newcommand{\fluxc}{$10^{-14}$ erg s$^{-1}$ cm$^{-2}$}
\newcommand{\fluxa}{$10^{-12}$ erg s$^{-1}$ cm$^{-2}$}

\newcommand{\vel}{\rm km s$^{-1}$}

\newcommand{\oxygen}{[O~{\sc iii}]}
\newcommand{\NII}{[N~{\sc ii}]~6584~\AA}

\newcommand{\OIII}{[O {\sc iii}]~5007~\AA}

\newcommand{\Ha}{H$\alpha$}

\title[Discovery of an optical cocoon tail behind the runaway HD 185806]{Discovery of an optical cocoon tail behind the runaway HD~185806}

\author[Spetsieri et al.]{Z. T. Spetsieri,$^{1,2}$\thanks{zoi.spetsieri@epfl.ch} P. Boumis,$^{2\orcidA{}}$\thanks{ptb@astro.noa.gr}
A. Chiotellis,$^{2\orcidA{}}$ S. Akras,$^{2\orcidC{}}$ S. Derlopa,$^{2}$ S. Shetye,$^{1}$
D. M.-A.~Meyer,$^{3}$
\newauthor{D. M. Bowman,$^{4\orcidD{}}$ V. V. Gvaramadze$^{5,6,7}$\thanks{dedicated to V.G. who passed away on 2 Sept. 2021}}
\\
 $^{1}$ Institute of Physics, Laboratory of Astrophysics, École Polytechnique Fédérale de Lausanne (EPFL), Observatoire de Sauverny, 1290
Versoix, Switzerland\\
$^{2}$ Institute for Astronomy, Astrophysics, Space Applications
and Remote Sensing, National Observatory of Athens,
15236 Penteli, Greece\\
$^{3}$ Universit\" at Potsdam, Institut f\" ur Physik und Astronomie, Karl-Liebknecht-Strasse 24/25, 14476 Potsdam, Germany\\
$^{4}$ Institute of Astronomy, KU Leuven, Celestijnenlaan 200D, B-3001 Leuven, Belgium\\
$^{5}$ Sternberg Astronomical Institute, Lomonosov Moscow State University, Universitetskij Pr. 13, Moscow 119992, Russia\\
$^{6}$ Space Research Institute, Russian Academy of Sciences, Profsoyuznaya 84/32, Moscow 117997, Russia\\
$^{7}$ E. Kharadze Georgian National Astrophysical Observatory, Abastumani 0301, Georgia\\
}

\date{Accepted 2022 June 29. Received 2022 June 24; in original form 2022 May 26}

\pubyear{2022}

\begin{document}
\label{firstpage}
\pagerange{\pageref{firstpage}--\pageref{lastpage}}
\maketitle

\begin{abstract}

Studies on the circumstellar structures around evolved stars provide vital information on the evolution of the parent star and the properties of the local interstellar medium. In this work, we present the discovery and characterization of an optical cocoon tail behind the star HD~185806. 

The cocoon apex emission is puzzling, as it is detected in the infrared but shows no signal in the optical wavelength.
The \ha~and \oxygen~ fluxes of the nebular structure vary from 2.7 to 8.5$\times$\fluxa\ and from 0.9 to 7.0$\times$\fluxb, respectively. 
Through high-resolution spectroscopy, we derive the spectral type of the star, construct the position-velocity diagrams of the cocoon tail for the \ha, \oxygen\ and \nii\ emission lines, and determine its velocity in the range of $-$100 to 40 \vel. Furthermore, we use SED fitting and MESA evolutionary models adopting a distance of 900~pc, and classify HD 185806  as a 1.3 M$_{\odot}$ star, in the transition phase between the RGB and early AGB stages. Finally, we study the morpho-kinematic structure of the cocoon tail using the astronomical software SHAPE. An ellipsoidal structure, with an inclination of $\sim19$\degr~with respect to the plane of sky is found to better reproduce the observed cocoon tail of HD 185806.
\end{abstract}




\section{Introduction}

 Several kinds of stars throughout the Hertzsprung-Russell (HR) diagram are characterized by intense mass outflows in the form of stellar winds. These winds drive gas and dust into the interstellar medium (ISM) and shape extended circumstellar structures around the parent star \citep{Langer2012,Smith2014}. Detailed studies of these circumstellar structures provide crucial insights on the nature and evolution of the mass-losing star, the properties of the local ISM and help us determine the impact of stellar outflows on the chemical and dynamical evolution of their host galaxies. 

 Wind shaped circumstellar bubbles have been found to display a variety of morphologies that deviate from the spherical symmetry being elliptical, bipolar, multipolar and revealing arcs, lobes, blobs, jets, hydrodynamical instabilities. Binary interactions \citep[]{DeMarco2009,vanMarle2012, Staff2016, Decin2019}, magnetic field activity \citep{Garcia-Segura1997,Garcia2005}, stellar rotation \citep[]{Hoefner1996,GarciaSegura1999, Garcia2016} are only some of the drivers of the morphologies formed around mass-losing stars. In addition, facts related to the ambient medium such as density gradients/discontinuities, magnetic fields, surrounding knots and clouds, pre-existing circumstellar structures etc., further contribute on the final outcome of the resulted circumstellar structures \citep[e.g.][]{Dwarkadas1996, Raga1998, Wareing2007, Esquivel2010, vanMarle2015}.

In the case where the mass-losing star is moving supersonically, with respect to the local ambient medium, a bow-shaped circumstellar structure is formed resulting by the balance of the wind's material and the ISM’s ram pressures \citep{Comeron1998}. Hence, bow-shaped wind structures are unique laboratories for determining or at least constraining the mass loss history of the central star, the surrounding ISM density and finally the stellar systemic motion \citep[e.g.][]{Huthoff2002,Gvaramadze2014,Meyer2014, Chiotellis2016,Meyer2016,meyer_mnras_496_2020}. 

Two main mechanisms have been proposed to be responsible for the formation of circumstellar structures around supersonically moving stars whose space velocities exceed the 30 $\rm km~s^{-1}$ a value that roughly corresponds to three times the typical stellar space velocity \citep[e.g.][]{Blaauw1956,Gies1986,Huthoff2002}. The first mechanism implies close dynamical interactions of either single stars \citep{Poveda1967}, binary-binary interactions \citep{Spitzer1980} or finally by triple body encounter between a binary and a single star \citep{Gvaramadze2011} that occur in a dense stellar environment such as in globular clusters. The second mechanism suggests that the fast moving star is a member of a binary system, which remained bound after the supernova explosion of the companion star \citep{Zwicky1957,Blaauw1961}. These two mechanisms are more likely to occur in massive stars. Indeed, the vast majority of bow shocks have been observed in massive OB runaway stars. In these stars, the dust and gas are accumulated behind the bow shock and are heated by the ultraviolet radiation of the massive star \citep[e.g.][]{Henney2019a}. Consequently, the dust radiates at the infrared band \citep{Draine1984}, while the shocked/photoionised gas produces optical line emission, such as \ha\ and \oiii. For this reason, bow shocks are mostly detected in the mid-infrared \citep[e.g.][]{vanBuren1995,Noriega-Crespo1997,Peri2012,Peri2015, Kobulnicky2016, Kobulnicky2017} and -less frequently- in the optical band \citep[e.g.][]{Gull1979,Kaper1997,Brown2005, Boumis2009}. In a few cases, non-thermal emission of bow shocks has also been detected in the radio (syncrotron) and X-ray (inverse Compton) bands \citep[e.g. BD+43$^o$3654, AE Aurigal;][respectively]{Benaglia2010, Lopez-Santiago2012}.

In the low stellar mass regime, nebular structures are mostly observed around stars in the red giant branch (RGB) and asymptotic giant branch (AGB) as during this phase substantial mass outflows are met \citep[e.g.][]{CDS,Weigelt2002, Schoier2006, Castro2010, Olofsson2010, Ali2013,Aguirre2020}. Most of these bow shocks have been detected in the mid-infrared band. A large Herschel/PACS survey of circumstellar structures around AGB stars revealed that bow shocks are a common feature met in about 40$\%$ of the sample \citep{Cox2012}. The stagnation point -i.e.  nearest distance between the central star and the circumstellar structure- was found to be smaller than 1~pc in all cases. Such a result agrees with the mass loss history expected from AGB stars, adopting typical Galactic warm neutral and ionised gas properties. 

\par This paper focuses on the circumstellar structure around the M-type, long period variable star \citep[]{HDvar2,HDvar3, HDvariability} at the constellation of Aquila, HD~185806 also known as V1279 Aql. This star had been observed with the Wide-field Infrared Survey Explorer \citep[WISE;][]{Wright2010} and displays a peculiar circumstellar structure that resembles a "tail" feature. However no further investigation has been conducted to identify the nature of the star or the properties of the surrounding nebula.

\par In this work, we present the discovery of an optical cocoon tail behind of HD 185806. We obtain narrow band imaging in several filters and high resolution optical spectroscopy, carrying out the first morpho-kinematic analysis of this cocoon tail in the optical regime. The paper is structured as follows: In Section~\ref{Observations}, we present the optical images and high dispersion spectra of the cocoon tail of HD 185806. The photometric and spectroscopic results of our analysis are shown in Section~\ref{Results}. An extended discussion regarding the properties of the star and its coccoon-tail, as well as, a 3D morphokinematical model of the circumstellar structure are presented in Section~\ref{Discussion}. We end with our conclusions in Section~\ref{Conclusions}.

\begin{table*}  
\caption{Observing log. Columns from left to right specify the filter, the central wavelength $\lambda_{\rm c}$, the wavelength difference ($\Delta \lambda$), the exposure time, the Observatory details and the date of the observations.}
\label{table1}
\begin{tabular}{ccccccc}  
\noalign{\smallskip}  
\hline  
\multicolumn{7}{c}{HIGH-RESOLUTION IMAGING with Aristarchos} \\  
\hline
Filter & $\lambda_{\rm c}$ & $\Delta \lambda$ & Exposure time & Observatory & Date & \\
  & (\AA) & (\AA) & (s)& & & \\
\hline
\ha\ + [N {\sc ii}] 6584 \AA & 6578 & 40 & 1800 & 2.3m Aristarchos & 2014 Aug 25-27 & \\
\ha\ 6563 \AA & 6567  & 17 & 1800 & 2.3m Aristarchos & 2014 Aug 25-27 & \\
\NII & 6588 & 17 & 1800 & 2.3m Aristarchos & 2014 Aug 25-27 & \\
\OIII & 5011 & 30 & 1800 & 2.3m Aristarchos & 2014 Aug 25-27 & \\
\hline

\multicolumn{7}{c}{HIGH-RESOLUTION SPECTROSCOPY with  HERMES} \\  
\hline
V & 3800–9000& -& 2000  & 1.2m Mercator Telescope & 2021 July 29 \\

\hline

\multicolumn{7}{c}{HIGH-RESOLUTION SPECTROSCOPY with MES} \\  
\hline  
Area & \multicolumn{2}{c}{Slit centres} & Exp. time & Slit P.A./ width & Observatory & Date \\  
 & $\alpha$ & $\delta$ & (sec) & (degrees / $\mu$m)& &\\  
\hline  
Slit 1 & 19\h40\m44.2\s & 02\degr30\arcmin09\arcsec & 1800 & 30 / 300 & 2.1m SPM & 2014 Nov 27\\
Slit 2 & 19\h40\m41.8\s & 02\degr30\arcmin57\arcsec & 1800 & 57 / 300 & 2.1m SPM & 2014 Nov 29\\
\hline

\hline  
\end{tabular}
\end{table*}

\section{Observations}
\label{Observations}
\subsection{High-resolution imaging}
Wide-field optical images of HD 185806 and its neighborhood were obtained with the 2.3m Aristarchos telescope at Helmos Observatory in Greece in 2014. The whole cocoon was imaged with four 5$\times$5 arcmin$^{2}$~pointings. A CCD detector with 1024$\times$1024 pixels$^{2}$~with 24$\mu$m pixel size, resulting to 0.28 arcsec pixel$^{-1}$~ resolution, was used for the observations. Exposures of 1800 s duration were obtained through the \ha, \hnii\ and \nii\ emission line filters on the main cocoon area, while the full mosaic was taken only with the \oiii\ filter \citep{Boumis2016}. In the latter, all fields were projected on to a common origin on the sky and were subsequently combined to create the final mosaic. The analysis of the images was carried out using the standard {\sc IRAF} routines for the bias subtraction, flat fielding correction, bad pixel and column masking, sky background estimation and cosmic-rays cleaning. The flux and astrometric calibration for all data frames was performed using the STScI Guide Star Catalogue~II \citep{Lasker2008}. We performed photometry of all sources appearing in the images although we were mainly focused on HD 185806. This was done in order to calibrate the magnitude and flux of the star based on a standard zero point and have a larger sample of stars to estimate the photometric errors. The photometric signal to noise ratio (S/N) of our images was S/N$>$15 and the chi square value $\chi{^2}\sim$2.0. 

\begin{figure*}
\centering
\includegraphics[scale=0.75]{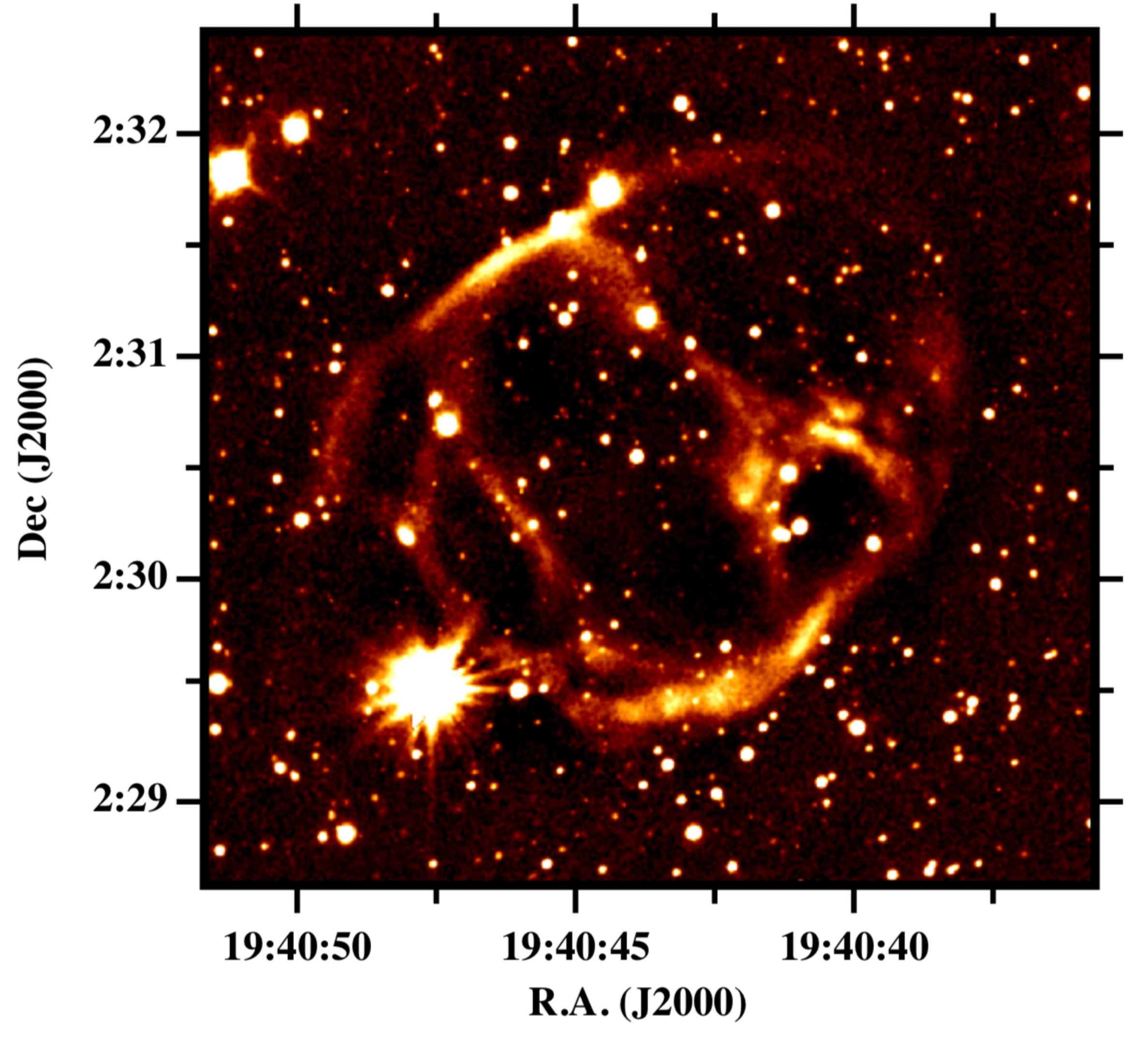}
\caption{Mosaic image of the region surrounding HD 185806 in \oiii\ emission. Shadings run linearly from 0 to 8.5$\times$\flux.}
\label{fig1}
\end{figure*}

\begin{figure*}
\centering
\includegraphics[scale=0.4]{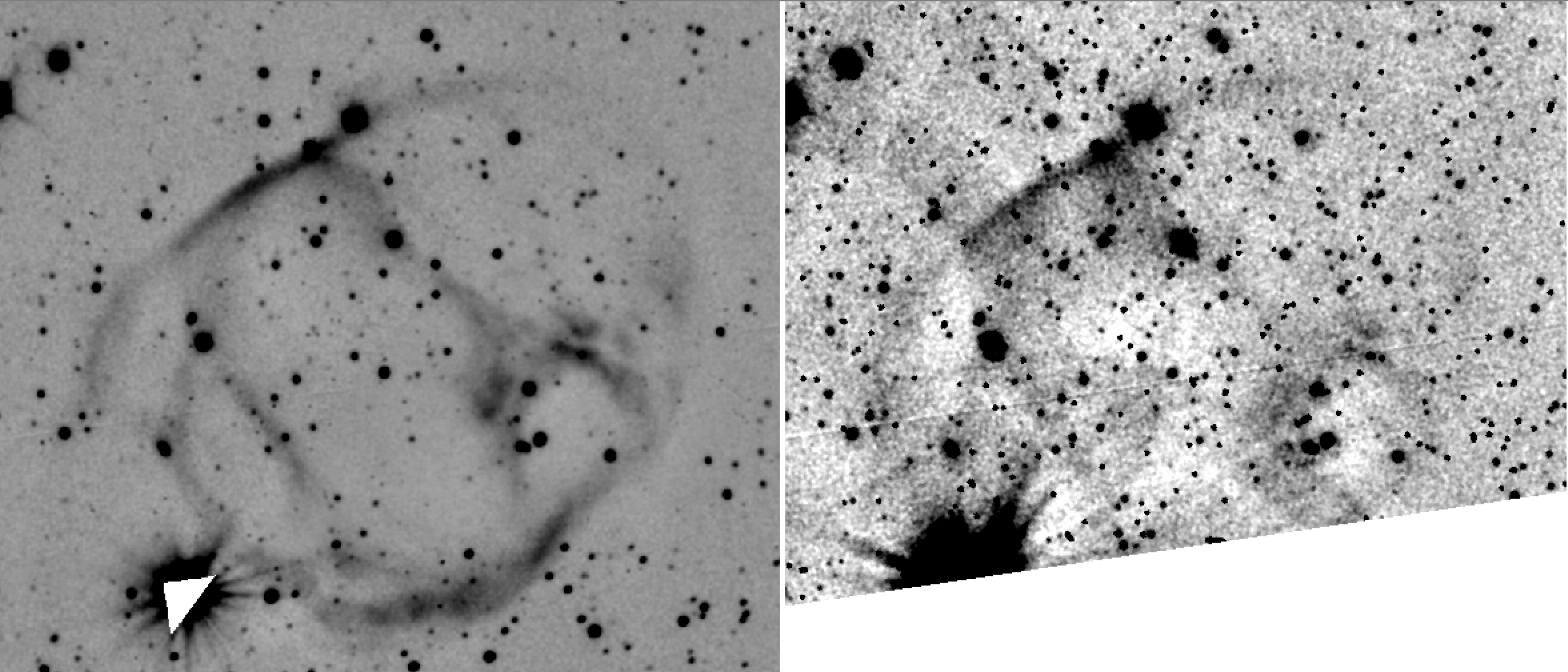}
\caption{A comparison between the \oiii\ (left) and the \ha\ (right) images of the same region of HD 185806. The faint ring structures in the \ha\ image are ghosts due to the brightness of the star. Shadings run linearly from 0 to 8.5$\times$\flux\ and 0 to 7.5$\times$\fluxb, respectively. The images scale is identical to that of Fig.~\ref{fig1}. North is to the top, East to the left.}
\label{fig2}
\end{figure*}

The resulting mosaic of the \oiii\ image, where HD 185806 and the whole cocoon tail can be seen is in Fig. \ref{fig1}, while in Fig. \ref{fig2} we show the comparison between the \oiii\ and the \ha\ emission line images. The \hnii\ and \nii\ images are not shown here, since the \hnii\ and \ha\ images are almost identical, with only a very few small regions of difference, where there is faint \nii\ emission. 

\subsection{High-dispersion and high resolution long-slit spectroscopy} 
High-dispersion long-slit echelle spectra were obtained at the 2.1~m telescope in the Observatorio Astron\'{o}mico of San Pedro Martir in Baja California, Mexico, in its {\it f}/7.5 configuration using the Manchester Echelle Spectrometer \citep{Meaburn2003}. The observations were carried out on 2014 November 27 and 29 (Table~\ref{table1}). 

\begin{figure*}
\centering
\includegraphics[scale=0.35]{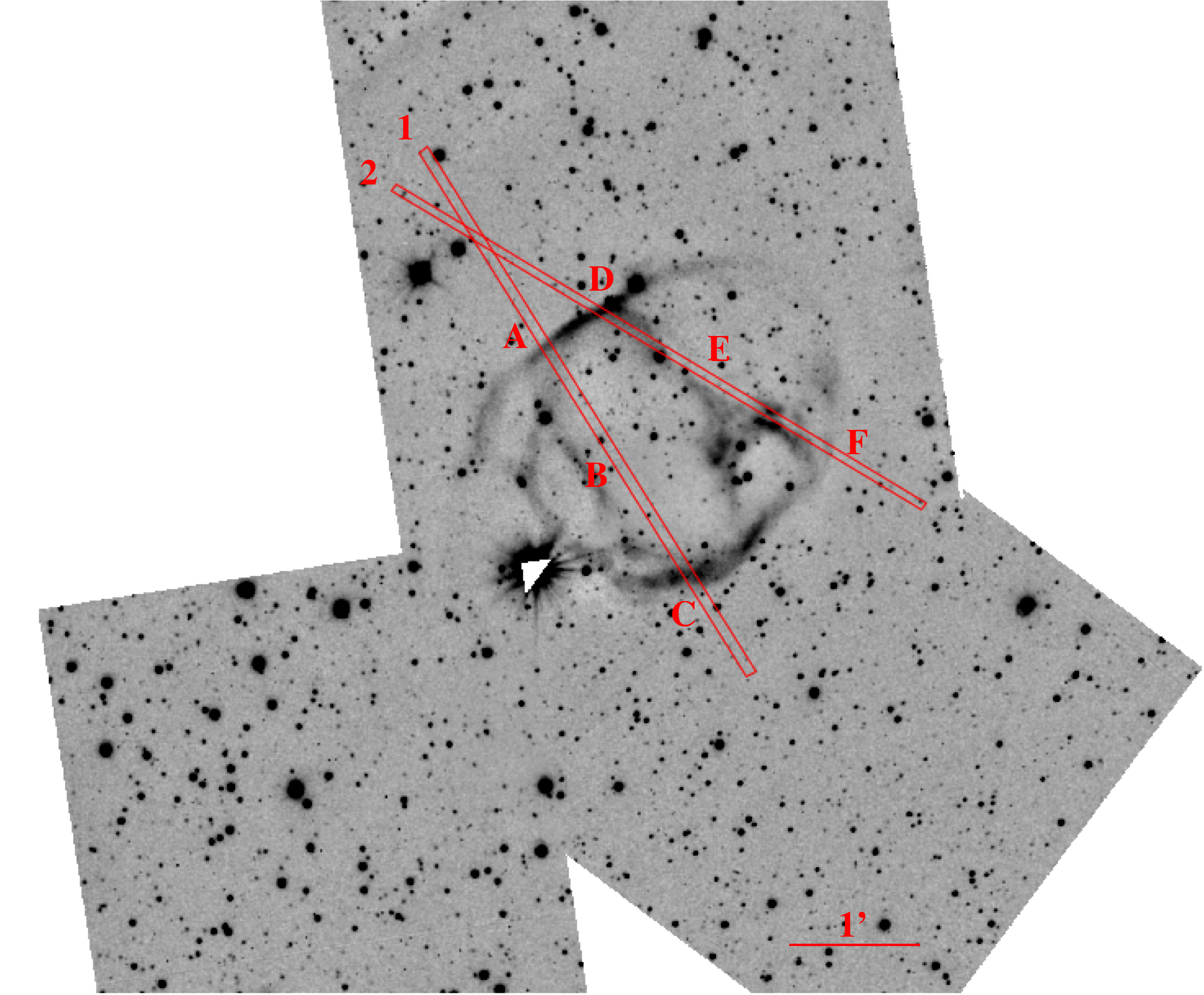}
\caption{Full mosaic image of HD 185806 in \oiii\ emission with slit positions 1 and 2 (in red). Shadings run linearly from 0 to 8.5$\times$\flux. North is to the top, East to the left.}
\label{fig3}
\end{figure*}

\begin{figure*}
\centering
\includegraphics[scale=0.5]{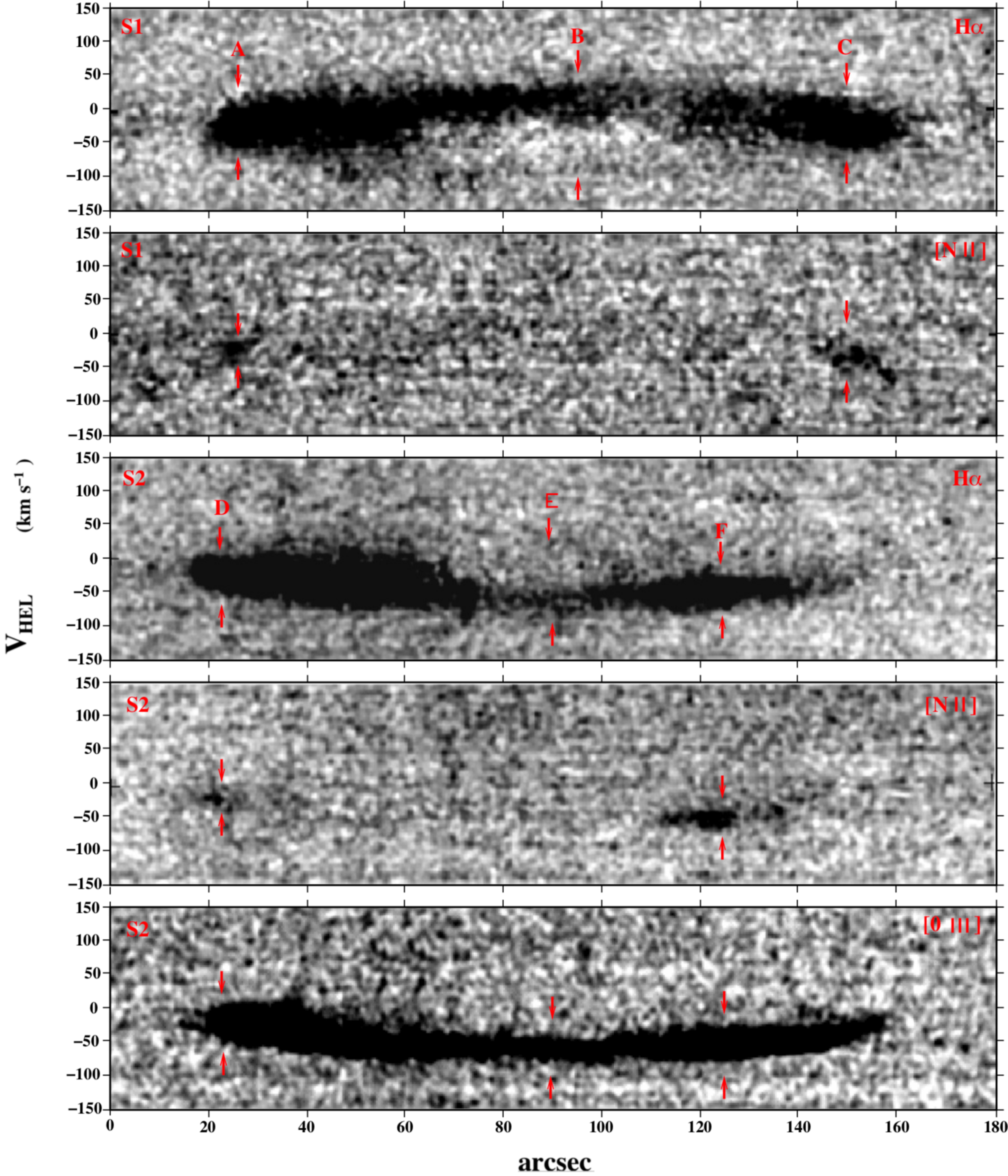}
\caption{Position-velocity images of slit positions 1-2 in the emission light of \ha, \nii\ and \oiii.}
\label{fig4}
\end{figure*}

A 300 $\mu$m ($\equiv$ 3.9 arcsec and 20 \vel) wide slit was used, orientated at position angles of 30\degr~and 57\degr. The 512 increments of the 2$\times$2 binned SITE CCD detector, each 0.624 arcsec long, give a total projected slit length of 5.32 arcmin on the sky. In this spectroscopic mode MES--SPM has no cross--dispersion consequently, for the present observations, a filter of 90-\AA\ bandwidth was used to isolate the 87$^{\rm th}$ echelle order containing the \ha\ and \niiab\ nebular emission lines, and a filter of 60-\AA\ bandwidth to isolate the 114$^{\rm th}$ echelle order containing the \oiii\ line. Integrations of 1800~s were obtained respectively for slit positions 1 and 2 marked in Fig. \ref{fig3}. The position--velocity (pv) arrays of \ha, \nii\ and \oiii\ line profiles from slit positions 1 and 2 are shown in Fig. \ref{fig4}. 

We conducted similar pre-processing steps as for the images such as bias subtraction, flat-fielding and illumination correction, cosmic-ray cleaning and background subtraction. The background subtraction in our case was conducted with extreme caution as over-subtraction of the background in filters with fainter emission would lead to erroneous assumptions about the spectral features of the bow shock structure. For the wavelength calibration of the spectra the Th-Ar arc-lamp  \citep{Meaburn2003} was used.
\par We additionally collected a high resolution spectrum of HD~185806 with the high-resolution fibre-fed HERMES spectrograph mounted on the 1.2-m Mercator Telescope at the Roque de Los Muchachos Observatory, La Palma (Canary Islands; \citealt{Raskin2011}). HERMES spectra have a spectral resolution of R = 85\,000 and a wavelength coverage 380–900~nm. The spectrum was automatically reduced through a dedicated pipeline at the end of the observing night. The observations were obtained in the V band with a signal-to-noise ratio of 100 to ensure the accurate determination of the stellar parameters and of the chemical abundances.
A summary and log of our imaging and spectral observations is provided in Table~\ref{table1}.

\begin{figure*}
\includegraphics[scale=0.35, trim=3cm 0 0 -5cm]{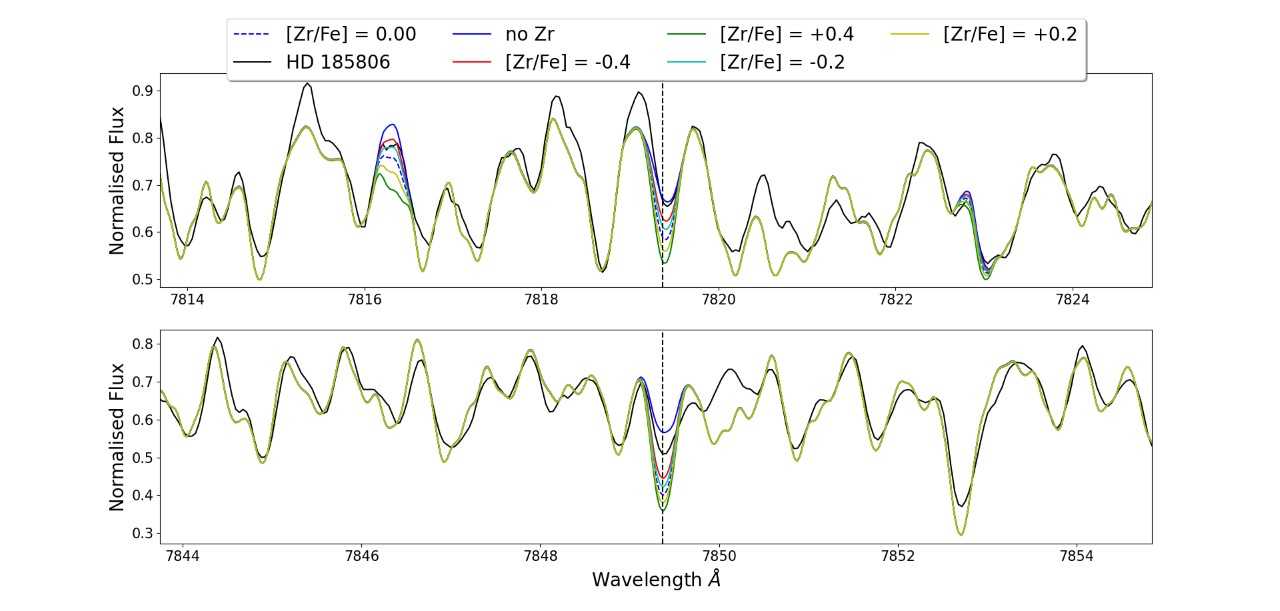}
\caption{Illustration of the quality of the match between observed and synthetic spectra obtained for HD 185806 around the Zr line at 7849.37\AA. The upper panel presents a $\pm$ 3 \AA zoom. The non-detection of Zr confirms that the star does not show any s-process enrichment.}
\label{fig8}
\end{figure*}

\begin{figure*}
\centering
\includegraphics[scale=0.75]{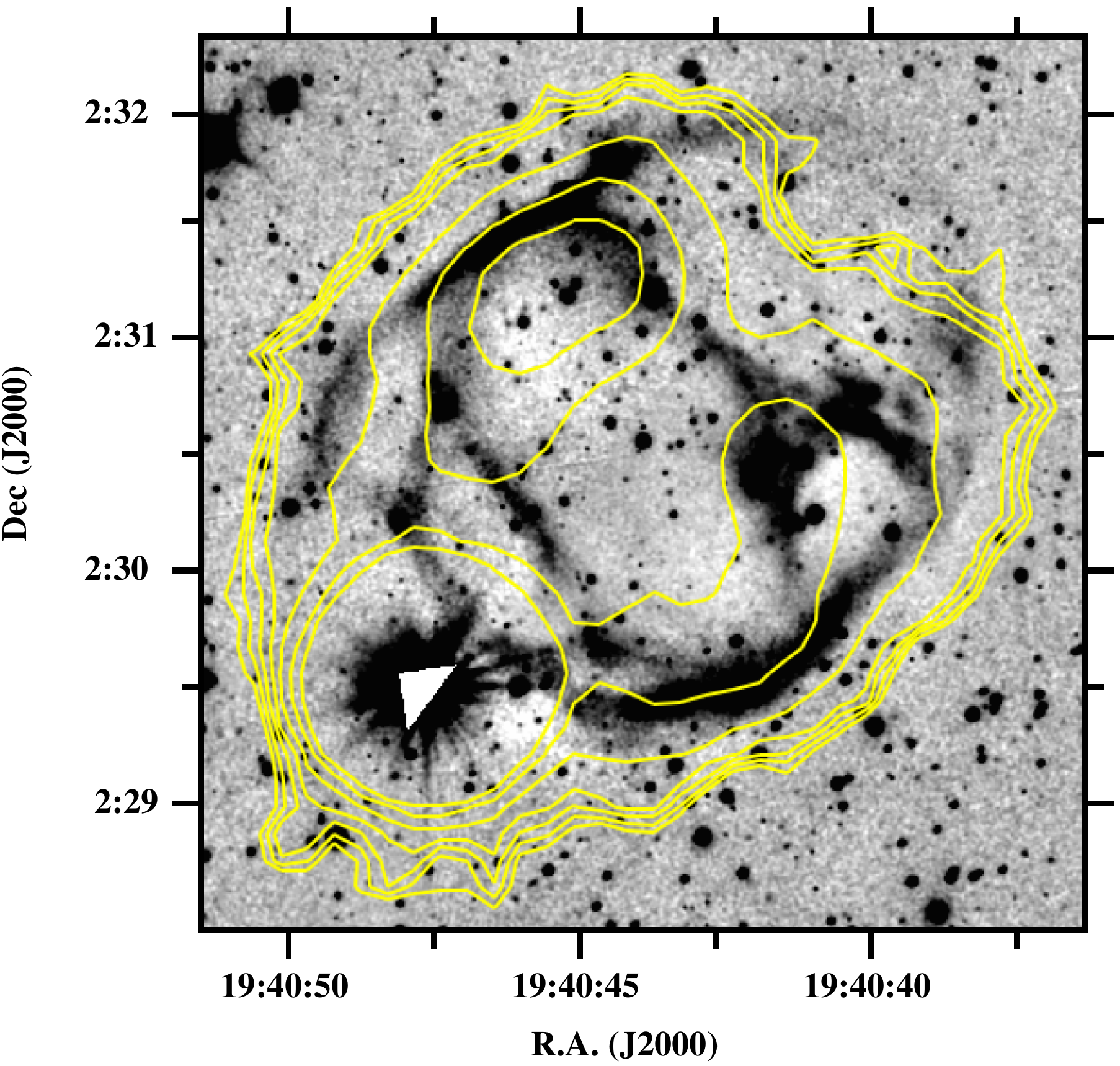}
\caption{The correlation between the \oiii\ and the W4 band at 22$\mu$m (yellow lines). The WISE contours produced by a combined image from the three different wavebands, where the star is bright in the first two and the nebula to the third. The contours scale linearly from 1$\times 10^{-5}$~counts arcmin$^{-2}$ s$^{-1}$ to 3$\times 10^{-4}$~counts arcmin$^{-2}$ s$^{-1}$, with step 2.8$\times 10^{-5}$~counts arcmin$^{-2}$ s$^{-1}$.}
\label{fig5}
\end{figure*}

\section{Results}
\label{Results}

\begin{table}
\begin{center}
\caption[]{Blue and red-shifted radial velocities in different positions across the slits, as shown in Fig.~\ref{fig3}. In the last column are the minimum and maximum velocities measured in each slit for the different emission lines.}
\label{table2}
\begin{tabular}{ccclc}
\hline
Slit & Position & Emission line & V$_{\rm hel}$ & V$_{\rm hel}$ $_{\sc min-max}$ \\
& & ($\rm \AA$) & (\,km\,sec$^{-1}$) & (\,km\,sec$^{-1}$) \\
\hline
1 & A & H$\alpha$ & $-75$ to $20$ & \\   
   & B &    & $-100$ to $40$ & $-100$ to $40$\\   
   & C &  & $-85$ to $20$ & \\     
\hline
1 & A & [N {\sc ii}] 6584 & $-50$ to $5$ & \\   
   & B &    &  & $-75$ to $35$\\   
   & C &  & $-65$ to $-7$ & \\     
\hline
2 & D & H$\alpha$ & $-55$ to $10$ & \\   
   & E &    & $-85$ to $30$ & $-100$ to $30$\\   
   & F &  & $-90$ to $-20$ & \\     
\hline
2 & D & [N {\sc ii}] 6584 & $-50$ to $5$ & \\   
   & E &    & & $-80$ to $5$\\   
   & F &  & $-80$ to $-30$ & \\     
 \hline  
2 & D & [O {\sc iii}] 5007 & $-55$ to $15$ & \\   
   & E &    & $-80$ to $-40$ & $-90$ to $15$\\   
   & F &  & $-60$ to $-20$ & \\     
\hline
\end{tabular}
\end{center}
\end{table}

\subsection{The optical emission line images}
Figure~\ref{fig1} displays the optical \oiii\ image of HD 185806 surrounded by an elliptical shape nebula, with a filamentary structure developed. The brighter filaments have \oxygen~fluxes from 7.1 to 8.5$\times$\flux\ and the fainter ones from 2.7 to 3.1$\times$\flux. For comparison, the \oiii\ and \ha~ emission of the nebula is shown in Fig.~\ref{fig2}. The nebula is evident in \ha~ displaying the same structure as in \oiii, however all filaments are fainter at this wavelength. The bright filaments have \ha~fluxes between 3.5 and 7.0$\times$\fluxb\ and the fainter ones between 8.5$\times$\fluxc\ and 1.5$\times$\fluxb. The \oxygen/\ha~flux ratio of the nebula is in the range from 10 to 30 being higher in the fainter filaments.

It is interesting to note that the emission is stronger in the opposite outer sides of the cocoon (areas A, C and D of Fig.~\ref{fig3} ) between $\alpha \simeq$ 19\h40\m45\s, $\delta \simeq$ 02\degr31\arcmin30\arcsec; $\alpha \simeq$ 19\h40\m47\s, $\delta \simeq$ 02\degr31\arcmin10\arcsec\ and $\alpha \simeq$ 19\h40\m40\s, $\delta \simeq$
02\degr30\arcmin00\arcsec; $\alpha \simeq$ 19\h40\m44\s, $\delta \simeq$ 02\degr29\arcmin25\arcsec. There are also two bright filaments almost perpendicular to the outer ones at $\alpha \simeq$ 19\h40\m46\s, $\delta \simeq$
02\degr30\arcmin15\arcsec\ (close to area B) and around $\alpha \simeq$ 19\h40\m42\s, $\delta \simeq$
02\degr30\arcmin30\arcsec\ (close to area E). On the contrary, both the bow-shock and the cocoon tail regions opposite to the star have either faint or no obvious emission.

\subsection{High-resolution spectra of the optical cocoon tail around HD 185806} 
 The full mosaic of HD 185806 with the cocoon tail surrounding it, is shown in Fig.~\ref{fig3}. The slit positions are indicated in red color while the regions where the heliocentric velocity of the gas was derived (Table \ref{table2}) are labeled as A, B, C, D, E, and F. Figure~\ref{fig4} shows the position velocity (PV) diagrams in the corresponding emission lines for slit 1 (S1) (top two panels) and slit 2 (S2) (lower three panels). A general result is that in all positions and in both slits, the whole structure of the cocoon reveals a bulk velocity towards us. Furthermore, at the same emission lines, there is a good agreement in velocities in positions from the same or nearby filaments, as it is for A and D, B and E, C and F. In Table~\ref{table2}, we also list the minimum and maximum velocities measured in each emission line, where it can be seen that the \ha\ ranges from $-$100 to 40 \vel, the \nii\ from $-$80 to 35 \vel\ and the \oiii\ from $-$90 to 15 \vel.  
 
 It is also noticeable in the PV diagrams that the \ha\ and \oiii\ emissions are detected all over the cocoon area, while in the \nii\ emission are detected only some bright spots (A/C and D/F for the S1 and S2 slit positions). The detection of the \ha, \nii, and \oiii\ emission lines in the cocoon tail of HD~185806 in conjunction with the low temperature of the star implies the presence of a shock-heated gas. If the orientation of the cocoon is taken into account, the expansion velocity of the ionized gas would be $>$ 70-80~\vel\ which can explain the strong \oiii\ emission. However, follow-up low-dispersion spectroscopic data are required in order to provide insights on the physical conditions of the gas.

\subsection{High-resolution spectrum of HD 185806}
\par The high resolution spectrum from HERMES yielded that HD~185806 is an M4-type, long period variable star. We estimated the atomic abundances following the same method as described in \citet{S18}. The spectra of S and M stars are dominated by molecular bands making the detection of unblended atomic s-process lines quite rare \citep{Smith1990}. This is why a spectral-synthesis approach is required, as opposed to relying solely on equivalent widths. 
In order to check whether the star is undergoing active slow neutron capture nucleosynthesis (s-process), which is typical in thermally-pulsing AGB stars, we checked for any presence of Technetium (Tc). Tc is an s-process element with no stable isotope. One of the isotopes 99Tc of the star has a half-life of 210\,000 yrs. This peculiar property of Tc makes it the best indicator for any active s-process nucleosynthesis in S and M type stars \citep{S18}. We checked the three Tc I lines located at 4239.191 \AA, 4267.27 \AA{} and 4297.06 \AA, and did not detect any absorption of Tc. We conclude that HD 185806 is not a thermally-pulsing AGB star. 
As in \citet{S18}, we used the two Zr lines at 7819.37 \AA{} and 7849.37 \AA{} with transition probabilities from laboratory measurements \citep{Biemont1981}. In the case of HD 185806 we do not report any overabundance of Zr, confirming that the star does not show any enrichment of s-process elements. We derived the [Fe/H] using some Fe lines at [Fe/H]= -0.50 dex revealing the sub-solar metallicity of the star. 
We additionally derived the nitrogen abundance from the CN lines in the 8000-8100 \AA range. In particular, using the lines listed in \citet{Merle2016}. We report a nitrogen abundance [N/Fe] = +0.6 to +1 dex.
An illustration of the observed and synthetic spectra obtained for HD~185806 around the Zr line at 7849.37 \AA{} are shown in Fig.~\ref{fig8}.

\begin{table*}  
\caption[]{Stellar properties of HD 185806.}  
\label{table3}
\begin{tabular}{ccccccccc}  
\noalign{\smallskip}  
\hline  
\hline
RA & Dec & T$_{\rm eff}$ & Luminosity & mass & parallax & radial velocity& transverse velocity & systemic velocity \\
(J2000)&(J2000)& K & L$_{\odot}$ & M$_{\odot}$ & mas& \vel& \vel& \vel\\
\hline
19:40:47.56 & +02:29:28.64 & 3400$\pm$100 & 1548 $\pm$200 & 1.3 & 1.109 &14.0& 43.75 & 45.0\\
\hline
\end{tabular}
\end{table*}  

\subsection{WISE observations}
HD~185806 and its cocoon tail have also been observed with WISE. The stellar source is detected in all bands (W1 (3.4~$\mu$m), W2 (4.6~$\mu$m), W3 (12.1~$\mu$m), and W4 (22~$\mu$m) \citep{Wright2010} but it is saturated in the first two. The cocoon tail is detected only in the W4 band which indicates the presence of warm dust.  
Fig.~\ref{fig5} shows the \oiii\ image with the contours from the W4 image over plotted in yellow. The contours surround both the star and the nebula, as well as, the interior filaments of the nebula revealing a strong correlation between the optical and infrared emission. This correlation is not uncommon, since it has been found in other cases, such as in Wolf-Rayet star nebulae \citep{Toala2015}. The W4 image shows that the emission in this band is dominated by thermal continuum emission from dust spatially coincident with the thin nebular shell or most likely with the leading edge of the nebula. Intriguingly, the bow shock that is absent from the optical images of HD~185806, is likely detected in the W4 band as it can be seen from the outer contours (Fig.~\ref{fig5}). 

This optical/infrared discrepancy at the location of the forward shock 
of the cocoon may partly originate from the penetration of interstellar dust into the inner circumstellar 
medium of HD~185806. This engenders a shift between the location of the outer layer of the cocoon and the location where the dust accumulates, as shown in \citet{vanmarle_apj_734_2011} in the context of the cool runaway red supergiant star Betelgeuse.
Finally, another deviation between the optical and infrared  occurs at the edge of the cocoon tail, opposite to the star, where the faint  \oiii\ filament is not accompanied by IR emission.

\section{Discussion}
\label{Discussion}
\subsection{Properties and nature of HD 185806}\label{Sect:4.1}
Our spectroscopic analysis confirmed that HD 185806 is an M4-type long period variable star. Based on Gaia eDR3 we estimated the distance of HD 185806 at D=909~pc \citep[]{Gaia2016,Gaia2018}. The angular velocity of the star is observed to be $\mu \sim 10.2~\rm mas~yr^{-1}$ and for the given distance corresponds to a transverse stellar velocity of $V_T = 43.75~\rm km~s^{-1}$. Finally, the Gaia eDR3 radial velocity value is found to be $V_r = 14.0~\rm km~s^{-1}$.

\begin{figure}
\includegraphics[scale=0.27]{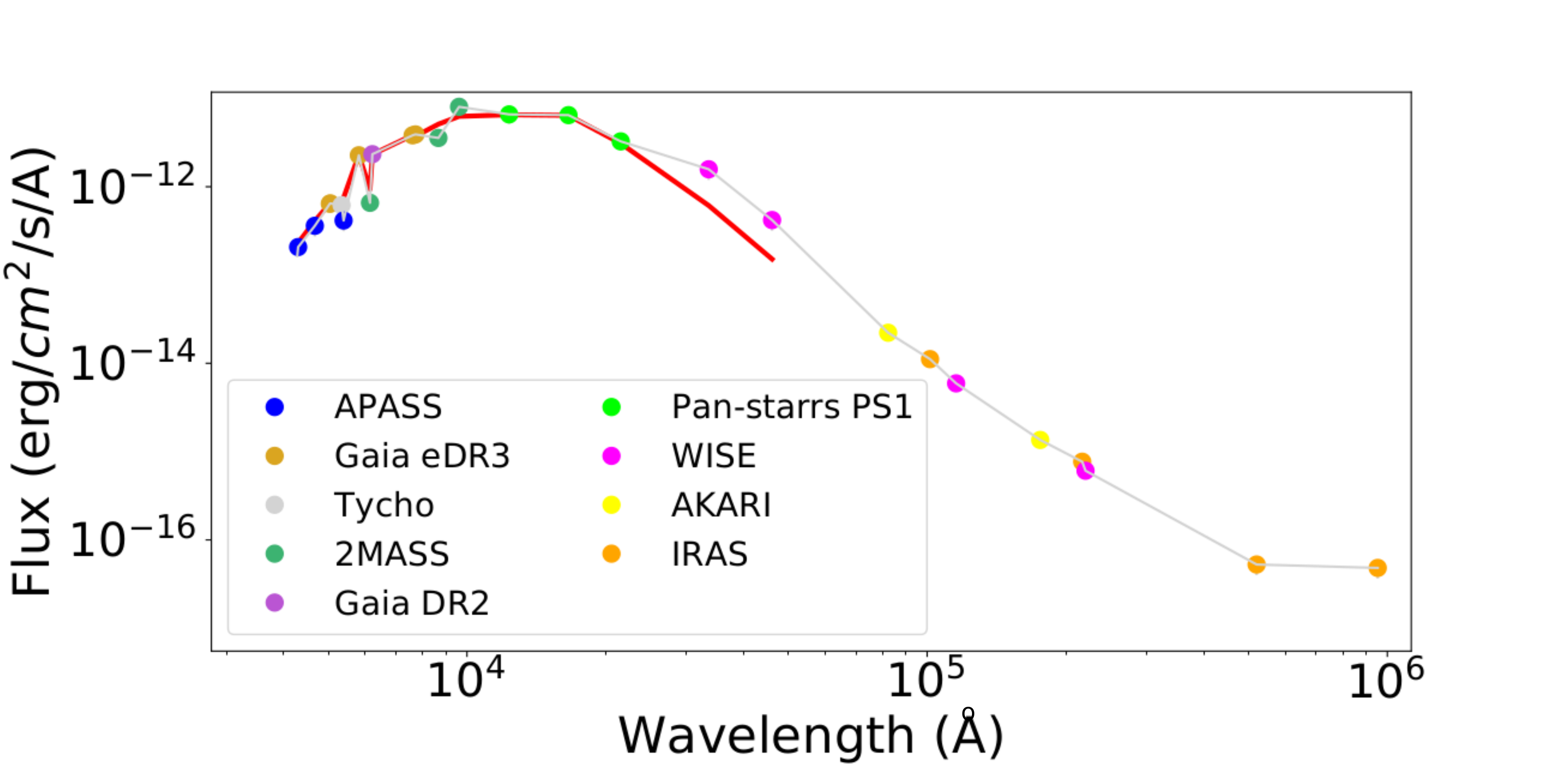}
\caption{Spectral Energy Distribution derived by VOSA. Colour-coded are the various photometric points from different surveys. Observations from 2MASS \citep{2MASS2006} are shown in green, from IRAS \citep{IRAS1984} in orange, from the AKARI \citep{AKARI2010} in yellow, from WISE \citep{Wright2010} in pink, from Tycho \citep{Tycho2000} in grey, from APASS \citep{APASS2016} in purple, from Gaia eDR3 \citep[]{Gaia2016,Gaia2018} in magenta, from Gaia eDR3 \citep{Gaiadr3} in brown andhttps://www.overleaf.com/project/6048838257fb47049f965521 Pan-Starrs PS1 \citep{PANSTARRS2017} in light green. The red curve is the SED fitting generated by high-resolution synthetic stellar spectra by \citep{Coelho2005}}.
\label{fig6}
\end{figure}

We used the photometric data available in virtual observatory spectral energy distribution (SED) analyzer \citep[VOSA;][]{VOSA} to determine the effective temperature (T$_{\rm eff}$) and luminosity (L) of the star. VOSA offers tools for the SED analysis, allowing the estimation of the stellar parameters. The synthetic photometry used by VOSA is calculated by convolving the response curve of the used filter set with the theoretical synthetic spectra. Then, a statistical test is performed, via $\chi{^2}$ minimization, to estimate which set of synthetic photometry best reproduces the observed data. 
The stellar parameters are obtained from the comparison of the observed spectrum with a simulated one with known stellar parameters. In our case, using VOSA we created the SED for HD 185806 and fitted the data with a library of high-resolution synthetic stellar spectra \citep[]{Coelho2005} created by the PFANT code \citep[]{Cayrel1991,Barbuy2018}. The SED generated for a distance of 900 pc  is shown in Fig.~\ref{fig6} and corresponds to T$_{\rm eff}$= 3400$\pm$100 K and L=1548$\pm$200 L$_{\odot}$. The parameters derived from VOSA seem to be in well agreement with the spectra. The extracted properties of HD 185806 are summarized in Table ~\ref{table3}.

\begin{figure}
\centering

\includegraphics[scale=0.29]{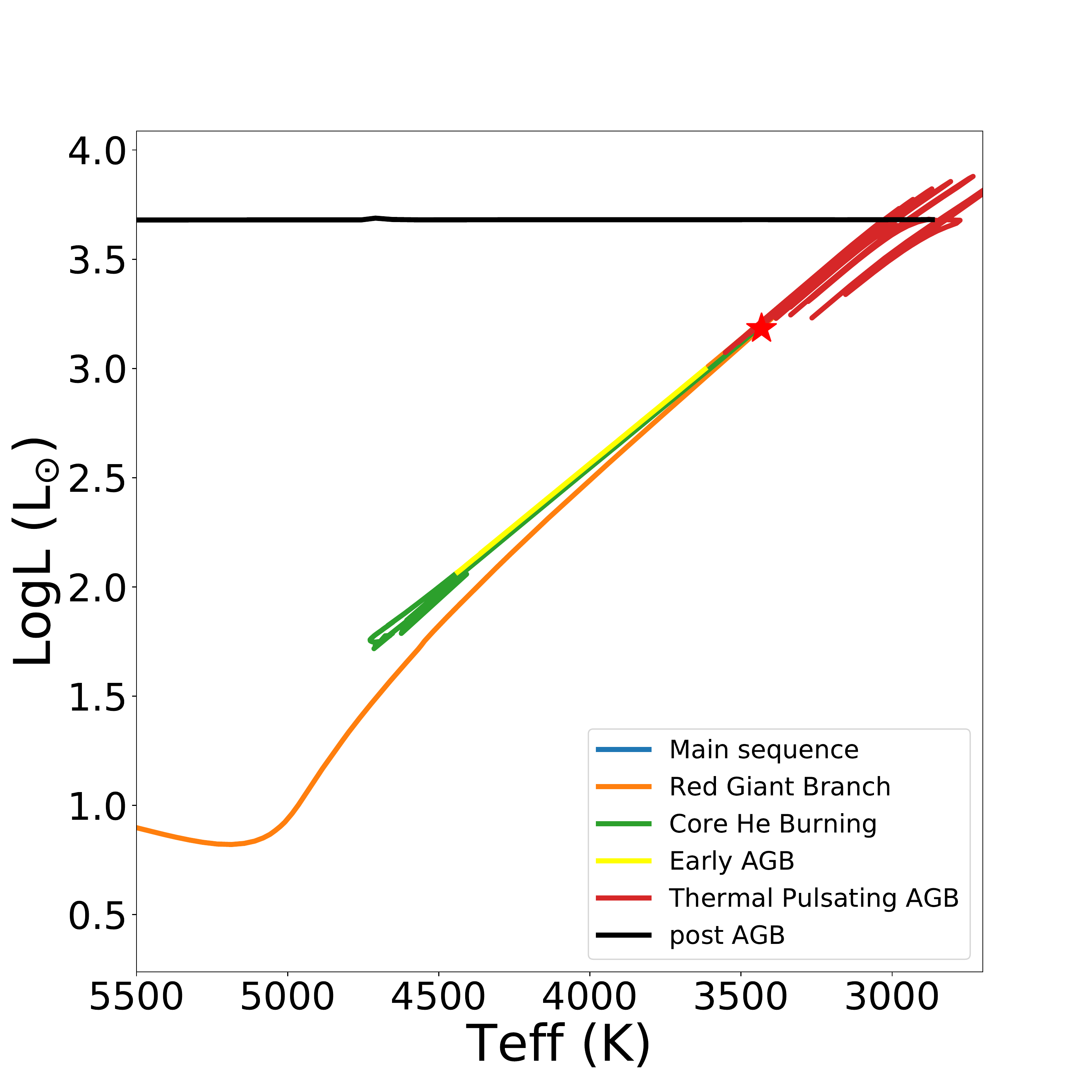}
\caption{Hertzsprung-Russell diagram generated from MESA for a 1.3 M$_{\odot}$ star. The phases of stellar evolution are displayed in different colors. The main sequence is shown in blue color, the red giant branch (RGB) is in orange, the Core He burning phase is shown in green, the early AGB phase in yellow, the thermal pulsating AGB phase is in red and the post AGB phase is in black color. HD 185806 is shown as a red star.}
\label{fig7}
\end{figure}

We employed the Modules for Experiments in Stellar Astrophysics (MESA) Isochrones and Stellar Tracks \citep[MIST;][]{MESA2016} to estimate the mass, age, and evolutionary state of HD 185806 considering solar metallicity. MESA models use solar-scaled abundances covering a wide range of ages, masses, phases, and metallicities computed within a single framework. The models were generated assuming rotation at 40 per cent of the critical breakup speed. Based on the spectral type, effective temperature, luminosity and [Fe/H], we plotted evolutionary tracks of 1-3 M$_{\odot}$ with steps of 0.1 M$_{\odot}$, using a Galactic extinction of A$_{V}$=1.12 mag by NASA/IPAC extra-galactic database \citep[NED;][]{NED}. The evolutionary track which best fits the stellar parameters of HD 185806 corresponds to an 1.3 M$_{\odot}$ star. Figure~\ref{fig7} illustrates the evolutionary track of a 1.3 M$_{\odot}$ star, in the HR diagram, together with the extracted parameters of HD 185806. The star is lying on the red giant branch (RGB) phase. This is consistent with the chemical signatures of the star that were discussed in Section \ref{Results}. Such an evidence is also aligned to the formation of the cocoon tail detected around HD 185806 as RGB stars are characterised by intense and slow stellar winds, able to shape dense circumstellar structures around the star.

\begin{figure}
\includegraphics[scale=0.55, trim=0cm 0 0 -5cm]{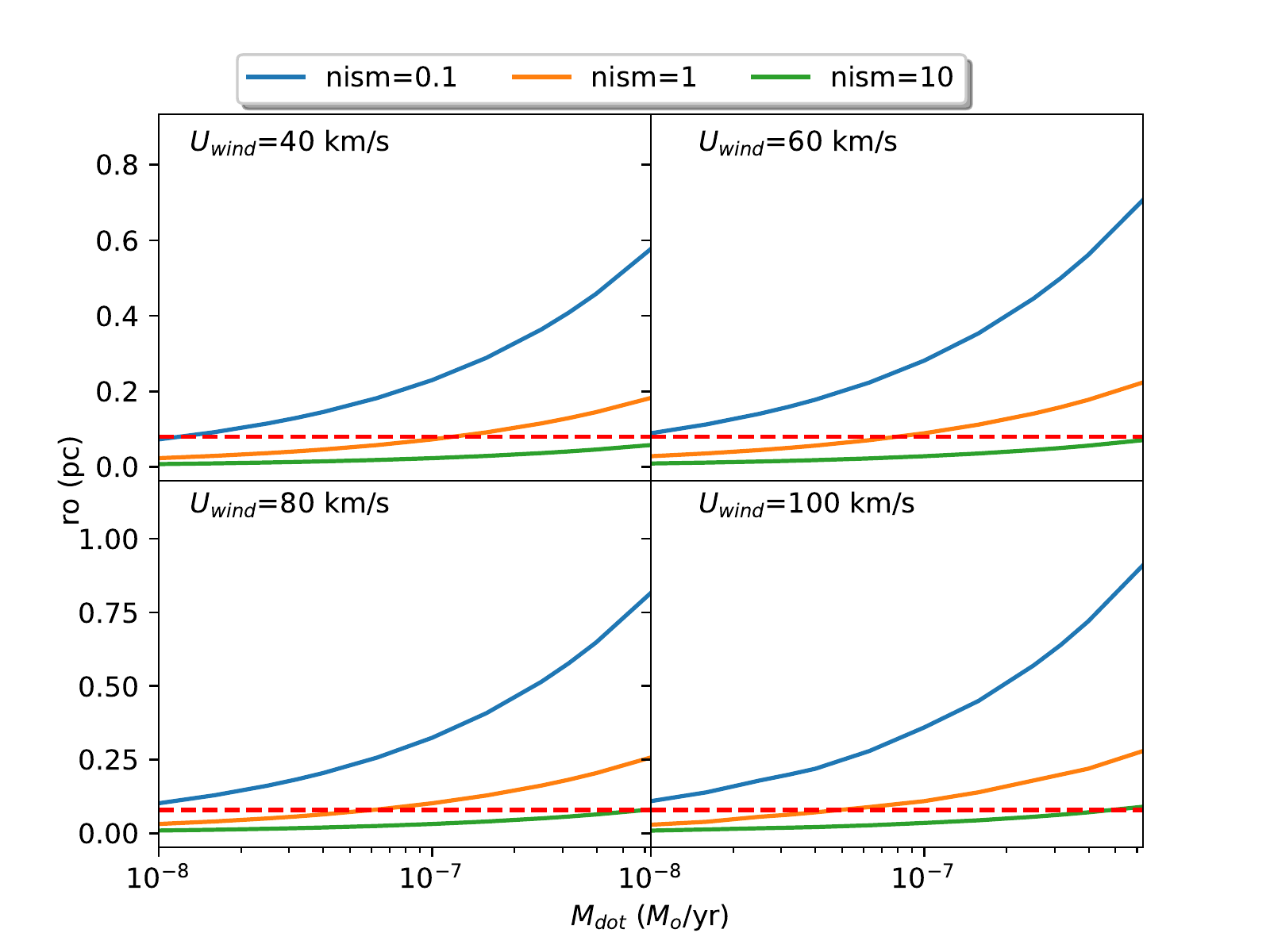}
\caption{Mass-loss vs. r$_{0}$ plot. The four panels show the distance of the bow shock from the star for varying values of wind velocity (U$_{wind}$) and ISM density (n$_{ism}$). The blue line corresponds to a n$_{ism}$=0.1, the orange line to n$_{ism}$=1 and the green line to n$_{ism}$=10. The dashed red line shows the distance below which the bow shock is not observable.}
\label{fig9}
\end{figure}

\subsection{On the morphology of the cocoon tail around HD 185806}

The evolutionary state and the  high spatial velocity of HD~185806 implies that the observed wind-blown bubble morphology around the star is dominantly determined by the interaction between the stellar wind material and the surrounding ISM. 

The circumstellar structures around supersonically moving stars are characterized by a bow-shaped shell formed in the direction of the stellar motion, followed by a cocoon/tail morphology. In the case where the wind and the ISM flows are supersonic, with respect to the sound speed of the local medium, shock waves are formed. The formed shock waves compress and heat up the medium that are propagating. The resulting bow shock structure consists of a forward shock separating the unshocked and shocked ISM, a wind termination shock that marks the border between the free-expanding wind and the shocked wind, and finally a contact discontinuity lies between the two shocks separating the shocked wind from the shocked ISM shells (see \citealt{Weaver1977} and \citealt{Lamers1999}). The position and shape of the bow shock is governed by the ram pressure balance between the stellar wind and the ISM, that in turn is determined by the wind mass loss rate (${\bm \dot{M}}$) and terminal velocity ($u_w$), the stellar systemic velocity ($u_*$) and the local ISM density ($\rho_{\rm ISM}$). In the direction of the stellar motion the bow shock is at its shorter distance - called as stagnation point - the radius of which is given by \citep{Baranov1971, Dyson1975, Wilkin1996}:

\begin{equation}
R_0 = \sqrt{\frac{{\bm \dot{M}}\ v_w}{4 \pi\ \rho_{ism}\ v^2_\star}} \label{eq:standoff}
\end{equation}
\\

The CSM around HD 185806 displays a cocoon-tail structure behind the star which indicates the presence of shocked ISM/wind material by the forward and termination shock of the wind bubble. However, even though a bow shock structure is clearly depicted at the WISE images  \citep{Wright2010} (see the yellow contours in Fig. \ref{fig5}), there is no evidence of the bow shock in the optical band. The current complex properties of the CSM around HD~185806 and the absence of an optically bright bow shock shell can be attributed to the following three main reasons (or to a combination of them): 

\subsubsection{Proximity of the bow shock to the central star}

In cases where the formed bow shock lies very close to the star it is likely to be out-shined by the stellar radiation or/and to be obscured by the produced stellar dust. To assess the plausibility of this statement in Fig. \ref{fig9} we illustrate the radial distance between the star and the stagnation point, as given by Eq.~(\ref{eq:standoff}), for different values of RGB stellar winds and of the ISM density. The dashed red line in Fig.~\ref{fig9} indicates the threshold below which the bow shock is not expected to be observed in the optical as it coincides with the area where the central star radiation dominates. As it is clearly portrayed, for the case where the ISM density is higher $n_{\rm ISM} \geq 1 cm^{-3}$~the predicted stagnation point radius is below the defined threshold for moderate mass loss rates ($\dot{M} 	\lesssim 10^{-7}~\rm M_{\odot}~yr^{-1}$), something that makes this scenario plausible. 

From a statistical point of view, half of the RGB/AGB stars and red supergiants with bow-shock shells studied by \cite{Cox2012} have a stagnation point at a de-projected distance of $<$0.08~pc, while there are 29 more stars without any detection. Therefore, the non-detection of a clear bow-shock in the HD 185806 is not unique or even rare, something that advocates further towards the credibility of the initial assumption.

\subsubsection{Time variable stellar wind properties} 

A second possible explanation of the non-detection of an optical emitting bow shock is based on the current evolutionary state of HD~185806. As we discussed in Section \ref{Sect:4.1}, the observed properties of the central star indicate that HD~185806 is currently at the RGB phase evolving towards the core helium burning/early AGB phase. If indeed the central star is in a transition state between two evolutionary stages it is likely that in the recent past its stellar wind properties (mass loss rate and wind velocity) have changed substantially. In particular, in this transition phase the mass loss rate is expected to decrease and the wind terminal velocity to increase. As a result, the freely expanding wind density drops significantly. In the framework of this scenario, the lack of an optically bright bow shock around HD~185806 is attributed to the low wind-bubble density in which the terminating shock is propagating. In addition, the information about the stellar wind properties alteration is not communicated spontaneously throughout the circumstellar structure but it travels within the dynamic timescale of the wind flow: $t_{\rm flow} \sim \frac{r} {u_w}$, where $r$ the distance from the mass-losing star and $u_w$ the wind velocity. Thus, there is the chance that we are currently witnessing a transition phase of the CSM around HD 185806, were in the region close to the stagnation point a low density wind is dominant, while the observed cocoon tail structure corresponds the imprint of an anterior phase where dense and bright CSM structures were formed.

\subsubsection{Deviation of the ambient medium from homogeneity} 

The above discussed arguments assume an uniform ambient medium in the warm phase ($T_{\rm ISM}\approx 1000-8000\, \rm K$) of the ISM of the Milky Way~\citep{mckee_80_aspc_1995}. The establishment of a stellar wind bow shock being both a function of the wind and ISM density combined with the absence of an observed arc around runaway stars, can indicate runaway stellar objects moving through a hot low-density region~\citep{huthoff_aa_383_2002}. However, this argument does not apply to HD 185806 since the bow shock is visible in near-infrared but not in optical. 

Nevertheless, any deviation from uniformity in the ambient medium affects the development of its circumstellar nebula. For example, the presence of native ISM overdensities and/or disturbances of the velocity field interacting with the forward shock of the bow shock induces distortions of the layers of shocked material that produce bright optical, free-free and near-infrared knots in the regions of shocked material \citep{baalmann_aa_650_2021,herbst_ssrv_218_2022}. Consequently, potential irregularities in the ISM density field in the vicinity of 
HD~185806 can only take the form of local underdensities which would make its bow shock deviate from the analytic solution of \citet{wilkin_459_apj_1996}. Such an example, is the production of a local extension of the forward shock, which is affected by Rayleigh-Taylor and other non-linear instabilities \citep{vishniac_apj_428_1994}. The star could be moving in an ISM gradually decreasing along the direction of stellar motion. This may explain the {\sc wise} $22\, \mu\rm m$ protuberances in the south-west region of Fig.~\ref{fig5}. 

Another assumption that has been tacitly made is that the ambient medium around HD~185806 is characterized by the local ISM properties. Thus, it has been neglected any ambient medium modifications that occurred in a prior phase of the central star’s evolution, namely the main-sequence (MS) phase. 

The complex internal morphology of the cocoon around HD~185806 could potentially be explained considering that the current circumstellar structure around HD~185806 resulted by the evolution of the RSG wind bubble within the interior of the previously formed MS bubble. The arcs and filamentary shapes visible via the \oiii\ emission of Fig.~\ref{fig5} not only trace the internal border of the cocoon, but also appear as bright structures parallel to the direction of stellar motion. The former can reasonably be associated to the contact discontinuity of the main-sequence wind-blown bubble at the moment it is shocked by the new-born RGB wind. Regarding the twisted optical arcs encompassed by the infrared cocoon we propose that they are projection effects similar to those found in the three-dimensional magneto-hydrodynamical simulations of~\citet{meyer_mnras_506_2021}. It is shown that the instabilities developing in the termination shock of the stellar wind bow shock of red supergiant stars can turn, by projection effects, into a large variety of optical H$\alpha$ filaments, including parallel optical segments to the direction of stellar motion, see Fig.~10f,h and Fig.~11e,h in ~\citet{meyer_mnras_506_2021}. We argue that a similar mechanism is at work inside of the cocoon of HD~185806. 

HD~185806 should have recently entered into its current, red giant branch, evolutionary phase, its circumstellar medium is not in a steady state~\citep{mohamed_aa_541_2012} and the time-variability of its stellar wind is a key parameter into the deep understanding of its cocoon. With this in mind it raises the question of its future evolution, particularly within the predictive models for fast-moving stars in the AGB, thermally-pulsating AGB stars and planetary nebula phase of~\citet{villaver_apj_632_2005}. The series of papers of \citet{villaver_apj_632_2005} strengthens the interpretation of the observed cocoon as a transient circumstellar structure generated by the onset of the RGB evolutionary phase, which might be replaced by an AGB wind to be released into the cocoon. The surroundings of HD~185806 may then eventually evolve to those of a thermal-pulsating AGB star, as depicted in the numerical models of~\citet{villalver_apj_748_2012}.

\begin{figure*}
\centering
\includegraphics[scale=0.4]{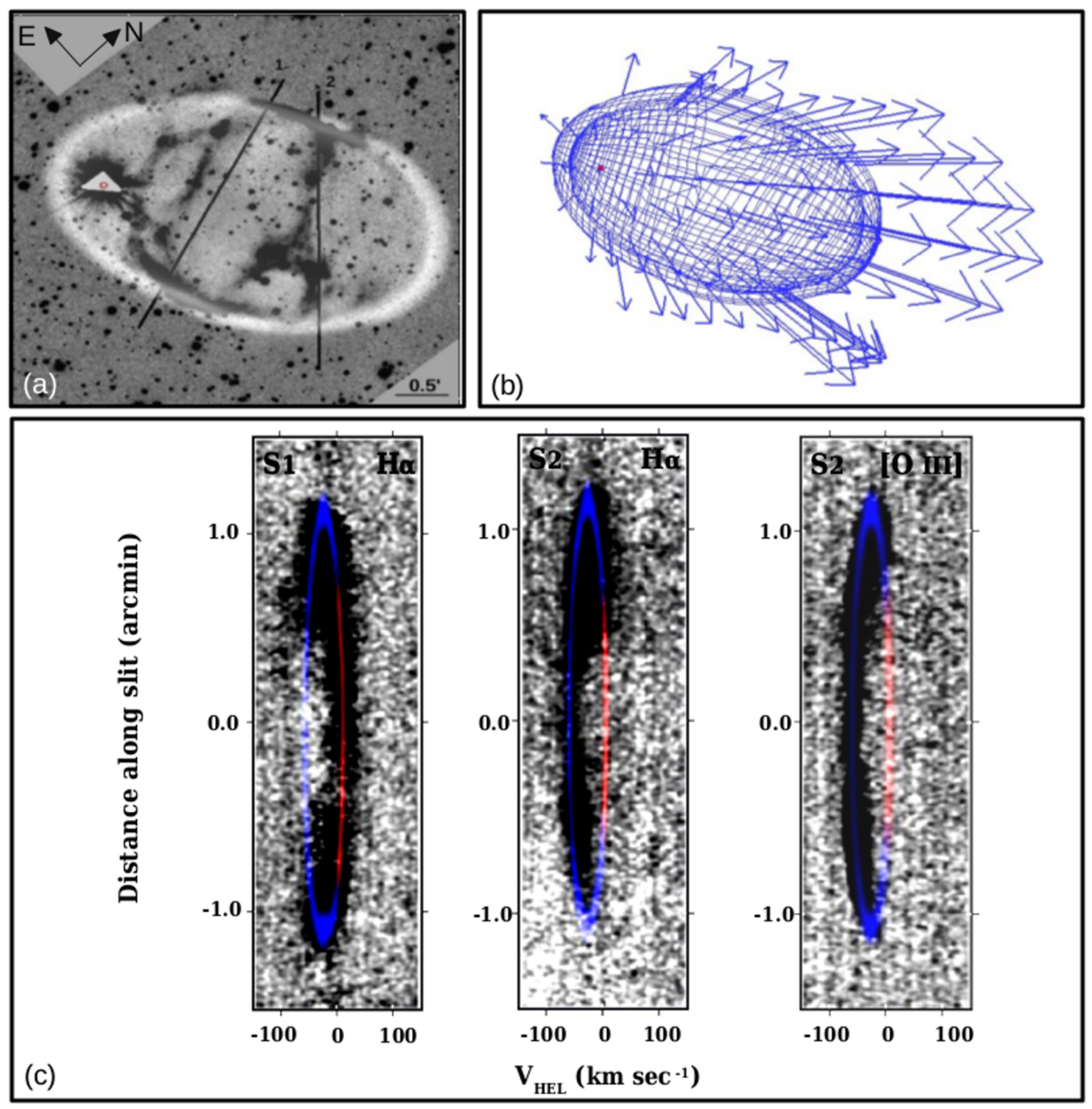}
\caption{(a) The 3D morpho-kinematic (MK) model of HD overlaid on its observational wide field image in \oiii. The two black lines represent the slits positions, while the red circle is the location of the star. (b) The 3D MK model in grid representation. The red dot is the location of the star, while the vectors are the velocity vectors that deduced from the velocity field of the shell of the cocoon tail. For description, see in the subsection \ref{3D MK}. (c) The observational PV diagrams in black deduced from the echelle spectra in \Ha ~and \oiii ~ emission lines, overlaid with the synthetic PV diagrams reproduced with SHAPE. Blue and red colours corresponds to the blue- and red-shifted gas.}
\label{fig10}
\end{figure*}

\subsection{Properties of the cocoon's forward shock}
The exact understanding of the morphology of the infrared/optical forward shocks of the cocoon around HD~185806 reveals a dynamical and internal synthesis of a nebula which exceeds that of a simple steady-state stellar wind bow shock. 
Further investigation would require more sophisticated numerical models. However, we can still elaborate on the shock properties by means of analytic arguments, as done in, e.g.~\citet{1992ApJ...386..265S,2011A&A...529A..14G, Chiotellis2012,Chiotellis2016,2017MNRAS.464.3229M}. 
In order to characterise the observed shock, we will use the 
methodology described in ~\citet{2011A&A...529A..14G}, and the latter works which were designed to characterise simulated hydrodynamical shocks. 

Bow shocks are mainly characterised by two quantities, namely the dynamical 
$t_{\rm dyn} = r_{0} / v$ and cooling 
$t_{\rm cool} = 3 k_{\rm B} T_{\rm ps} / n_{\rm ps} \Lambda( T_{\rm ps} )$ timescales, 
respectively, where $k_{\rm B}$ is the Boltzmann constant, $T_{\rm ps}$ the post-shock 
temperature at the forward shock, $n_{\rm ps}$ the post-shock density at the forward shock 
and $\Lambda(T_{\rm ps})$ the corresponding cooling rate by optically-thin processes, 
see~\citet{Meyer2014}. 
According to Fig.~\ref{fig9}, the stand-off distance of the cocoon is $\approx 0.1 
\le r_{0} \le 1.0 \, \rm pc$. Since the bulk motion of the star is 
$v_{\star}=45\, \rm km\, \rm s^{-1}$, we have 
$v=v_{\star}/4 \approx 11.25\, \rm km\, \rm s^{-1}$ and therefore 
$t_{\rm dyn}\approx 0.0088$$-$$0.088\, \rm Myr$ assuming 
$r_{0}=0.1$$-$$1.0\, \rm pc$.  

Similarly, the cooling timescale is obtained by estimating first the conditions in the 
post-shock region at the forward shock using the relations in section 5.3 
of~\citet{2011A&A...529A..14G}. This gives $T_{\rm ps} \approx 1.4\, \times 10^{5}
( v_{\star} / 100\, \rm km\, \rm s^{-1} )^{2} \approx 28350\, \rm K$ and 
$n_{\rm ps} = 4 n_{\rm ISM}=0.4$$-$$40 \, \rm cm^{-3}$ (assuming the explored 
parameter space of Fig.~\ref{fig9}). Using the cooling curve detailed 
in~\citet{Meyer2014}, one can find $ \Lambda( T_{\rm ps} )=
4.78 \times 10^{-23}\, \rm erg\, \rm cm^{3}\, \rm s^{-1}$. 
Consequently, $t_{\rm cool}=0.019$$-$$0.00019\, \rm Myr$ for 
$n_{\rm ISM}=0.1$$-$$10 \, \rm cm^{-3}$. 

The creation of large bow shock ($r_{0}=1.0\, \rm pc$) is possible with a diluted ambient 
only ($n_{\rm ISM}=0.1\, \rm cm^{-3}$), for which it is found that 
$t_{\rm cool}=0.019  \sim  t_{\rm dyn}\approx 0.088\, \rm Myr$, 
indicating that the shock is adiabatic, with both timescales being 
of the same order of magnitude. 
The creation of small bow shock ($r_{0}=0.1\, \rm pc$) is possible in a dense 
ISM ($n_{\rm ISM}=10\, \rm cm^{-3}$), gives 
$t_{\rm cool}=0.00019  \ll  t_{\rm dyn}\approx 0.0088\, \rm Myr$, 
meaning that the shock is radiative. 

The stellar evolution calculations performed in the context of HD 185806 show 
that its mass-loss rate is $\dot{M}\le10^{-7}\, \rm M_{\odot}\, \rm yr^{-1}$ 
This value favors the scenario of the existence of a small bow shock ($r_{0}=0.1\, \rm pc$)
and therefore of a radiative forward shock of the circumstellar cocoon 
of HD 185806. 
This should additionally be confirmed by multi-dimensional hydrodynamical calculations in our following studies.

\subsection{3D morpho-kinematic modeling of the cocoon shell}
\label{3D MK}
For the Morpho-kinematic (MK) modeling of the cocoon tail of HD~185806 in three-dimensions, we employed the astronomical software SHAPE \citep{Steffen2011}. This code has been widely used to model morpho-kinematic structures of planetary nebulae \citep[e.g. ][]{akras2012a,Akras2012,Akras2015,Clyne2015}, nova shells \citep[]{Harvey2020} and more recently complex supernova remnants \citep{Derlopa2020}. Despite the cocoon tail of HD 185806 displays a simple ellipsoidal morphology, the intriguing formation of the nebula behind the highly moving star motivated us to reproduce the observed position-velocity diagram and reconstruct the 3D velocity field.

The wide field image of HD 185806 in the \oiii\ line, along with the PV diagrams in the \ha\ and \oiii\ emission lines were used to constrain the morpho-kinematic structure of the cocoon tail. An ellipsoidal structure with a projected size of 3.6$\times$2.4~arcmin$^2$~can adequately reproduce the nebula behind the HD 185806 star. The position angle of the cocoon tail is found to be $\sim$19\degr~and the inclination angle $\sim$10\degr, both with respect to the plane of the sky, i.e. the star goes inwards the page, as it is shown in Fig.~\ref{fig10}(a).

For the expansion velocity law of the shell gas, a homologous expansion law ($\overrightarrow{V}$ $=$ $k$ $\overrightarrow{r}$) was applied. The centre of the velocity law is on the HD 185806 star and not at the geometrical centre of the nebula such as in PNe, resulting in faster expanding gas at larger distances from the star. Fig.~\ref{fig10}(b) shows the 3D grid structure of the cocoon tail along with the vectors of the velocity field (magnitude and direction). The red dot represents the location of the star. Fig.~\ref{fig10}(c) illustrates the matching between the observed (black) and modelled (colored) PV diagrams reproduced with SHAPE. According to our model, the mean value for the $k$ constant parameter in the homologous expansion is $31.3$ \,km\,sec$^{-1}$ arcmin$^{-1}$.

\section{Conclusions}
\label{Conclusions}

We have investigated the properties of the source HD 185806 and of its newly detected optical cocoon tail. By performing a photometric and spectroscopic analysis of the source in the optical filters H$\alpha$, \oiii, H$\alpha$+\nii~and \nii~we report that:  

\begin{itemize}

\item{The star is an M type star followed by an optical cocoon tail with intense \oiii~emission. The \oiii~flux of the filaments ranges from 2.7 to 8.5$\times$\flux. The spectroscopic analysis of the results shows that the heliocentric expansion velocity of the filaments ranges between -100 to 40~\vel, and the systemic velocity is \\45~\vel.}

\item{We used the virtual observatory SED analyzer \citep[VOSA;][]{VOSA} in which we imported all the available photometric data of the star and constructed the SED of the source from 4000~\AA~ up to 100~$\micron$. The SED implies an effective temperature T$_{\rm eff}$=3400$\pm$100~K and logL=3.18~L$_{\odot}$ as well as an infrared excess. The stellar parameters obtained from VOSA were used to create an HR diagram using the MESA models and evolutionary tracks. MESA models implied that the star is a 1.3 M$_{\odot}$ star in the RGB/early-AGB phase.} 

\item{We investigated the absence of an optically bright bow shock shell around HD 185806 which can be attributed to two main factors: the distance of the stagnation point regarding the star and the current evolutionary state of HD 185806. In the first case the bow shock could be outshined by the stellar radiation or/and be obscured by the large amount of dust that lies close to the central star (Fig. \ref{fig9}). In the second case, as the models imply that the star is in the transition phase between the RGB/early AGB phase, the wind mass-loss rate undergoes a decrease and the wind terminal velocity increases. As a result, the freely expanding wind density props by several orders of magnitude and the dense stellar wind bubble is transformed into a low density cavity. Regarding the statistical facts provided in \cite{Cox2012} the non-detectability of a bow shock in the optical regime is not uncommon.}

\item{The 3D morpho-kinematic model we created for the cocoon tail shows that its morphology can be reproduced by an ellipsoidal structure with a projected size of 3.6$\times$2.4~arcmin$^2$, a position angle of $\sim$19\degr~and an inclination of $\sim$10\degr~(Fig.~\ref{fig10}a). We studied the expansion velocity using a homologous law ($\overrightarrow{V}$ $=$ $k$ $\overrightarrow{r}$) as shown in Fig.~\ref{fig10}(b) and, according to our model, the mean value for the $k$ constant parameter is $31.3$ \,km\,sec$^{-1}$ arcmin$^{-1}$}.
\end{itemize}

\section*{Acknowledgements}
This research is co-financed by Greece and the European Union (European Social Fund-ESF) through the Operational Programme “Human Resources Development, Education and Lifelong Learning 2014-2020” in the context of the project “On the interaction of Type Ia Supernovae with Planetary Nebulae” (MIS 5049922). A.C. acknowledge the support of this work by the project ``PROTEAS II'' (MIS 5002515),which is implemented under the Action ``Reinforcement of the Research and Innovation Infrastructure'', funded by the Operational Programme ``Competitiveness, Entrepreneur- ship and Innovation'' (NSRF 2014–2020) and co-financed by Greece and the European Union (European Regional Development Fund). SA acknowledges support under the grant 5077 financed by IAASARS/NOA. DMB gratefully acknowledges a senior postdoctoral fellowship from the Research Foundation Flanders (FWO) with grant agreement no. 1286521N.
This work is based on observations made with the ``Aristarchos'' telescope operated on the Helmos Observatory by the Institute of Astronomy, Astrophysics, Space Applications and Remote Sensing of the National Observatory of Athens. This work has made use of data from the European Space Agency (ESA) mission Gaia (https://www.cosmos.esa.int/gaia), processed by the Gaia Data Processing and Analysis Consortium (DPAC, https://www.cosmos.esa.int/web/gaia/dpac/consortium). Funding for the DPAC has been provided by national institutions, in particular the institutions participating in the Gaia Multilateral Agreement. Data products from the Wide-field Infrared Survey Explorer, which is a joint project of the University of California, Los Angeles, and the Jet Propulsion Laboratory/California Institute of Technology, funded by the National Aeronautics and Space Administration.
The authors acknowledge the North-German Supercomputing Alliance (HLRN) for providing HPC resources 
that have contributed to the research results reported in this paper. 
The authors would like to dedicate this article to Vasilii V. Gvaramadze who passed away on 2 September 2021, since we started this project together.
\section*{Data Availability}

The data underlying this article will be shared on reasonable request to the corresponding author.

\bibliographystyle{mnras}
\bibliography{bowshock}

\begin{thebibliography}{}
\makeatletter
\relax
\def\mn@urlcharsother{\let\do\@makeother \do\$\do\&\do\#\do\^\do\_\do\%\do\~}
\def\mn@doi{\begingroup\mn@urlcharsother \@ifnextchar [ {\mn@doi@}
  {\mn@doi@[]}}
\def\mn@doi@[#1]#2{\def\@tempa{#1}\ifx\@tempa\@empty \href
  {http://dx.doi.org/#2} {doi:#2}\else \href {http://dx.doi.org/#2} {#1}\fi
  \endgroup}
\def\mn@eprint#1#2{\mn@eprint@#1:#2::\@nil}
\def\mn@eprint@arXiv#1{\href {http://arxiv.org/abs/#1} {{\tt arXiv:#1}}}
\def\mn@eprint@dblp#1{\href {http://dblp.uni-trier.de/rec/bibtex/#1.xml}
  {dblp:#1}}
\def\mn@eprint@#1:#2:#3:#4\@nil{\def\@tempa {#1}\def\@tempb {#2}\def\@tempc
  {#3}\ifx \@tempc \@empty \let \@tempc \@tempb \let \@tempb \@tempa \fi \ifx
  \@tempb \@empty \def\@tempb {arXiv}\fi \@ifundefined
  {mn@eprint@\@tempb}{\@tempb:\@tempc}{\expandafter \expandafter \csname
  mn@eprint@\@tempb\endcsname \expandafter{\@tempc}}}

\bibitem[\protect\citeauthoryear{{Akras} \& {L{\'o}pez}}{{Akras} \&
  {L{\'o}pez}}{2012}]{Akras2012}
{Akras} S.,  {L{\'o}pez} J.~A.,  2012, \mn@doi [\mnras]
  {10.1111/j.1365-2966.2012.21578.x}, \href
  {https://ui.adsabs.harvard.edu/abs/2012MNRAS.425.2197A} {425, 2197}

\bibitem[\protect\citeauthoryear{{Akras} \& {Steffen}}{{Akras} \&
  {Steffen}}{2012}]{akras2012a}
{Akras} S.,  {Steffen} W.,  2012, \mn@doi [\mnras]
  {10.1111/j.1365-2966.2012.20928.x}, \href
  {https://ui.adsabs.harvard.edu/abs/2012MNRAS.423..925A} {423, 925}

\bibitem[\protect\citeauthoryear{{Akras}, {Boumis}, {Meaburn}, {Alikakos},
  {L{\'o}pez}  \& {Gon{\c{c}}alves}}{{Akras} et~al.}{2015}]{Akras2015}
{Akras} S.,  {Boumis} P.,  {Meaburn} J.,  {Alikakos} J.,  {L{\'o}pez} J.~A.,
  {Gon{\c{c}}alves} D.~R.,  2015, \mn@doi [\mnras] {10.1093/mnras/stv1468},
  \href {https://ui.adsabs.harvard.edu/abs/2015MNRAS.452.2911A} {452, 2911}

\bibitem[\protect\citeauthoryear{{Alfonso-Garz{\'o}n}, {Domingo}, {Mas-Hesse}
  \& {Gim{\'e}nez}}{{Alfonso-Garz{\'o}n} et~al.}{2012}]{HDvariability}
{Alfonso-Garz{\'o}n} J.,  {Domingo} A.,  {Mas-Hesse} J.~M.,   {Gim{\'e}nez} A.,
   2012, \mn@doi [\aap] {10.1051/0004-6361/201220095}, \href
  {https://ui.adsabs.harvard.edu/abs/2012A&A...548A..79A} {548, A79}

\bibitem[\protect\citeauthoryear{{Ali}, {Ismail}, {Snaid}  \& {Sabin}}{{Ali}
  et~al.}{2013}]{Ali2013}
{Ali} A.,  {Ismail} H.~A.,  {Snaid} S.,   {Sabin} L.,  2013, \mn@doi [\aap]
  {10.1051/0004-6361/201321798}, \href
  {https://ui.adsabs.harvard.edu/abs/2013A&A...558A..93A} {558, A93}

\bibitem[\protect\citeauthoryear{{Baalmann}, {Scherer}, {Kleimann}, {Fichtner},
  {Bomans}  \& {Weis}}{{Baalmann} et~al.}{2021}]{baalmann_aa_650_2021}
{Baalmann} L.~R.,  {Scherer} K.,  {Kleimann} J.,  {Fichtner} H.,  {Bomans}
  D.~J.,   {Weis} K.,  2021, \mn@doi [\aap] {10.1051/0004-6361/202039836},
  \href {https://ui.adsabs.harvard.edu/abs/2021A&A...650A..36B} {650, A36}

\bibitem[\protect\citeauthoryear{{Baranov}, {Krasnobaev}  \&
  {Kulikovskii}}{{Baranov} et~al.}{1971}]{Baranov1971}
{Baranov} V.~B.,  {Krasnobaev} K.~V.,   {Kulikovskii} A.~G.,  1971, Soviet
  Physics Doklady, \href
  {https://ui.adsabs.harvard.edu/abs/1971SPhD...15..791B} {15, 791}

\bibitem[\protect\citeauthoryear{{Barbuy}, {Trevisan}  \& {de
  Almeida}}{{Barbuy} et~al.}{2018}]{Barbuy2018}
{Barbuy} B.,  {Trevisan} J.,   {de Almeida} A.,  2018, {PFANT: Stellar spectral
  synthesis code} (\mn@eprint {ascl} {1812.003})

\bibitem[\protect\citeauthoryear{{Bayo}, {Rodrigo}, {Barrado Y Navascu{\'e}s},
  {Solano}, {Guti{\'e}rrez}, {Morales-Calder{\'o}n}  \& {Allard}}{{Bayo}
  et~al.}{2008}]{VOSA}
{Bayo} A.,  {Rodrigo} C.,  {Barrado Y Navascu{\'e}s} D.,  {Solano} E.,
  {Guti{\'e}rrez} R.,  {Morales-Calder{\'o}n} M.,   {Allard} F.,  2008, \mn@doi
  [\aap] {10.1051/0004-6361:200810395}, \href
  {https://ui.adsabs.harvard.edu/abs/2008A&A...492..277B} {492, 277}

\bibitem[\protect\citeauthoryear{{Benaglia}, {Romero}, {Mart{\'\i}}, {Peri}  \&
  {Araudo}}{{Benaglia} et~al.}{2010}]{Benaglia2010}
{Benaglia} P.,  {Romero} G.~E.,  {Mart{\'\i}} J.,  {Peri} C.~S.,   {Araudo}
  A.~T.,  2010, \mn@doi [\aap] {10.1051/0004-6361/201015232}, \href
  {https://ui.adsabs.harvard.edu/abs/2010A&A...517L..10B} {517, L10}

\bibitem[\protect\citeauthoryear{{Biemont}, {Grevesse}, {Hannaford}  \&
  {Lowe}}{{Biemont} et~al.}{1981}]{Biemont1981}
{Biemont} E.,  {Grevesse} N.,  {Hannaford} P.,   {Lowe} R.~M.,  1981, \mn@doi
  [\apj] {10.1086/159213}, \href
  {https://ui.adsabs.harvard.edu/abs/1981ApJ...248..867B} {248, 867}

\bibitem[\protect\citeauthoryear{{Blaauw}}{{Blaauw}}{1956}]{Blaauw1956}
{Blaauw} A.,  1956, \mn@doi [\pasp] {10.1086/126983}, \href
  {https://ui.adsabs.harvard.edu/abs/1956PASP...68..495B} {68, 495}

\bibitem[\protect\citeauthoryear{{Blaauw}}{{Blaauw}}{1961}]{Blaauw1961}
{Blaauw} A.,  1961, \bain, \href
  {https://ui.adsabs.harvard.edu/abs/1961BAN....15..265B} {15, 265}

\bibitem[\protect\citeauthoryear{{Boumis}, {Meaburn}, {Lloyd}  \&
  {Akras}}{{Boumis} et~al.}{2009}]{Boumis2009}
{Boumis} P.,  {Meaburn} J.,  {Lloyd} M.,   {Akras} S.,  2009, \mn@doi [\mnras]
  {10.1111/j.1365-2966.2009.14784.x}, \href
  {https://ui.adsabs.harvard.edu/abs/2009MNRAS.396.1186B} {396, 1186}

\bibitem[\protect\citeauthoryear{{Boumis}, {Akras}, {Leonidaki}, {Chiotellis},
  {Kopsacheili}, {Alikakos}, {Nanouris}  \& {Mavromatakis}}{{Boumis}
  et~al.}{2016}]{Boumis2016}
{Boumis} P.,  {Akras} S.,  {Leonidaki} I.,  {Chiotellis} A.,  {Kopsacheili} M.,
   {Alikakos} J.,  {Nanouris} N.,   {Mavromatakis} F.,  2016, in Supernova
  Remnants: An Odyssey in Space after Stellar Death. p.~15

\bibitem[\protect\citeauthoryear{{Brown} \& {Bomans}}{{Brown} \&
  {Bomans}}{2005}]{Brown2005}
{Brown} D.,  {Bomans} D.~J.,  2005, \mn@doi [\aap]
  {10.1051/0004-6361:20041054}, \href
  {https://ui.adsabs.harvard.edu/abs/2005A&A...439..183B} {439, 183}

\bibitem[\protect\citeauthoryear{{Castro-Carrizo} et~al.,}{{Castro-Carrizo}
  et~al.}{2010}]{Castro2010}
{Castro-Carrizo} A.,  et~al., 2010, \mn@doi [\aap]
  {10.1051/0004-6361/201014755}, \href
  {https://ui.adsabs.harvard.edu/abs/2010A&A...523A..59C} {523, A59}

\bibitem[\protect\citeauthoryear{{Cayrel}, {Perrin}, {Buser}, {Barbuy}  \&
  {Coupry}}{{Cayrel} et~al.}{1991}]{Cayrel1991}
{Cayrel} R.,  {Perrin} M.~N.,  {Buser} R.,  {Barbuy} B.,   {Coupry} M.~F.,
  1991, \aap, \href {https://ui.adsabs.harvard.edu/abs/1991A&A...247..122C}
  {247, 122}

\bibitem[\protect\citeauthoryear{{Chambers} \& {et al.}}{{Chambers} \& {et
  al.}}{2017}]{PANSTARRS2017}
{Chambers} K.~C.,  {et al.} 2017, VizieR Online Data Catalog, \href
  {https://ui.adsabs.harvard.edu/abs/2017yCat.2349....0C} {p. II/349}

\bibitem[\protect\citeauthoryear{{Chiotellis}, {Schure}  \&
  {Vink}}{{Chiotellis} et~al.}{2012}]{Chiotellis2012}
{Chiotellis} A.,  {Schure} K.~M.,   {Vink} J.,  2012, \mn@doi [\aap]
  {10.1051/0004-6361/201014754}, \href
  {https://ui.adsabs.harvard.edu/abs/2012A&A...537A.139C} {537, A139}

\bibitem[\protect\citeauthoryear{{Chiotellis}, {Boumis}, {Nanouris}, {Meaburn}
  \& {Dimitriadis}}{{Chiotellis} et~al.}{2016}]{Chiotellis2016}
{Chiotellis} A.,  {Boumis} P.,  {Nanouris} N.,  {Meaburn} J.,   {Dimitriadis}
  G.,  2016, \mn@doi [\mnras] {10.1093/mnras/stv2798}, \href
  {https://ui.adsabs.harvard.edu/abs/2016MNRAS.457....9C} {457, 9}

\bibitem[\protect\citeauthoryear{{Choi}, {Dotter}, {Conroy}, {Cantiello},
  {Paxton}  \& {Johnson}}{{Choi} et~al.}{2016}]{MESA2016}
{Choi} J.,  {Dotter} A.,  {Conroy} C.,  {Cantiello} M.,  {Paxton} B.,
  {Johnson} B.~D.,  2016, \mn@doi [\apj] {10.3847/0004-637X/823/2/102}, \href
  {http://adsabs.harvard.edu/abs/2016ApJ...823..102C} {823, 102}

\bibitem[\protect\citeauthoryear{{Clyne}, {Akras}, {Steffen}, {Redman},
  {Gon{\c{c}}alves}  \& {Harvey}}{{Clyne} et~al.}{2015}]{Clyne2015}
{Clyne} N.,  {Akras} S.,  {Steffen} W.,  {Redman} M.~P.,  {Gon{\c{c}}alves}
  D.~R.,   {Harvey} E.,  2015, \mn@doi [\aap] {10.1051/0004-6361/201526585},
  \href {https://ui.adsabs.harvard.edu/abs/2015A&A...582A..60C} {582, A60}

\bibitem[\protect\citeauthoryear{{Coelho}, {Barbuy}, {Mel{\'e}ndez}, {Schiavon}
   \& {Castilho}}{{Coelho} et~al.}{2005}]{Coelho2005}
{Coelho} P.,  {Barbuy} B.,  {Mel{\'e}ndez} J.,  {Schiavon} R.~P.,   {Castilho}
  B.~V.,  2005, \mn@doi [\aap] {10.1051/0004-6361:20053511}, \href
  {https://ui.adsabs.harvard.edu/abs/2005A&A...443..735C} {443, 735}

\bibitem[\protect\citeauthoryear{{Comeron} \& {Kaper}}{{Comeron} \&
  {Kaper}}{1998}]{Comeron1998}
{Comeron} F.,  {Kaper} L.,  1998, \aap, \href
  {https://ui.adsabs.harvard.edu/abs/1998A&A...338..273C} {338, 273}

\bibitem[\protect\citeauthoryear{{Cox} et~al.,}{{Cox} et~al.}{2012}]{Cox2012}
{Cox} N.~L.~J.,  et~al., 2012, \mn@doi [\aap] {10.1051/0004-6361/201117910},
  \href {https://ui.adsabs.harvard.edu/abs/2012A&A...537A..35C} {537, A35}

\bibitem[\protect\citeauthoryear{{De Marco}}{{De Marco}}{2009}]{DeMarco2009}
{De Marco} O.,  2009, \mn@doi [\pasp] {10.1086/597765}, \href
  {https://ui.adsabs.harvard.edu/abs/2009PASP..121..316D} {121, 316}

\bibitem[\protect\citeauthoryear{{Decin}}{{Decin}}{2019}]{Decin2019}
{Decin} L.,  2019, in ALMA2019: Science Results and Cross-Facility Synergies.
  p.~53, \mn@doi{10.5281/zenodo.3585272}

\bibitem[\protect\citeauthoryear{{Derlopa}, {Boumis}, {Chiotellis}, {Steffen}
  \& {Akras}}{{Derlopa} et~al.}{2020}]{Derlopa2020}
{Derlopa} S.,  {Boumis} P.,  {Chiotellis} A.,  {Steffen} W.,   {Akras} S.,
  2020, \mn@doi [\mnras] {10.1093/mnras/staa2336}, \href
  {https://ui.adsabs.harvard.edu/abs/2020MNRAS.499.5410D} {499, 5410}

\bibitem[\protect\citeauthoryear{{Draine} \& {Lee}}{{Draine} \&
  {Lee}}{1984}]{Draine1984}
{Draine} B.~T.,  {Lee} H.~M.,  1984, \mn@doi [\apj] {10.1086/162480}, \href
  {https://ui.adsabs.harvard.edu/abs/1984ApJ...285...89D} {285, 89}

\bibitem[\protect\citeauthoryear{{Dwarkadas}, {Chevalier}  \&
  {Blondin}}{{Dwarkadas} et~al.}{1996}]{Dwarkadas1996}
{Dwarkadas} V.~V.,  {Chevalier} R.~A.,   {Blondin} J.~M.,  1996, \mn@doi [\apj]
  {10.1086/176772}, \href
  {https://ui.adsabs.harvard.edu/abs/1996ApJ...457..773D} {457, 773}

\bibitem[\protect\citeauthoryear{{Dyson}}{{Dyson}}{1975}]{Dyson1975}
{Dyson} J.~E.,  1975, \mn@doi [\apss] {10.1007/BF00636999}, \href
  {https://ui.adsabs.harvard.edu/abs/1975Ap&SS..35..299D} {35, 299}

\bibitem[\protect\citeauthoryear{{Esquivel}, {Raga}, {Cant{\'o}},
  {Rodr{\'\i}guez-Gonz{\'a}lez}, {L{\'o}pez-C{\'a}mara}, {Vel{\'a}zquez}  \&
  {De Colle}}{{Esquivel} et~al.}{2010}]{Esquivel2010}
{Esquivel} A.,  {Raga} A.~C.,  {Cant{\'o}} J.,  {Rodr{\'\i}guez-Gonz{\'a}lez}
  A.,  {L{\'o}pez-C{\'a}mara} D.,  {Vel{\'a}zquez} P.~F.,   {De Colle} F.,
  2010, \mn@doi [\apj] {10.1088/0004-637X/725/2/1466}, \href
  {https://ui.adsabs.harvard.edu/abs/2010ApJ...725.1466E} {725, 1466}

\bibitem[\protect\citeauthoryear{{Gaia Collaboration} et~al.,}{{Gaia
  Collaboration} et~al.}{2016}]{Gaia2016}
{Gaia Collaboration} et~al., 2016, \mn@doi [\aap]
  {10.1051/0004-6361/201629272}, \href
  {https://ui.adsabs.harvard.edu/abs/2016A&A...595A...1G} {595, A1}

\bibitem[\protect\citeauthoryear{{Gaia Collaboration} et~al.,}{{Gaia
  Collaboration} et~al.}{2018}]{Gaia2018}
{Gaia Collaboration} et~al., 2018, \mn@doi [\aap]
  {10.1051/0004-6361/201833051}, \href
  {https://ui.adsabs.harvard.edu/abs/2018A&A...616A...1G} {616, A1}

\bibitem[\protect\citeauthoryear{{Gaia Collaboration} et~al.,}{{Gaia
  Collaboration} et~al.}{2021}]{Gaiadr3}
{Gaia Collaboration} et~al., 2021, \mn@doi [\aap]
  {10.1051/0004-6361/202039657}, \href
  {https://ui.adsabs.harvard.edu/abs/2021A&A...649A...1G} {649, A1}

\bibitem[\protect\citeauthoryear{{Garcia-Segura}}{{Garcia-Segura}}{1997}]{Garcia-Segura1997}
{Garcia-Segura} G.,  1997, \mn@doi [\apjl] {10.1086/316796}, \href
  {https://ui.adsabs.harvard.edu/abs/1997ApJ...489L.189G} {489, L189}

\bibitem[\protect\citeauthoryear{{Garc{\'\i}a-Segura}, {Langer},
  {R{\'o}{\.z}yczka}  \& {Franco}}{{Garc{\'\i}a-Segura}
  et~al.}{1999}]{GarciaSegura1999}
{Garc{\'\i}a-Segura} G.,  {Langer} N.,  {R{\'o}{\.z}yczka} M.,   {Franco} J.,
  1999, \mn@doi [\apj] {10.1086/307205}, \href
  {https://ui.adsabs.harvard.edu/abs/1999ApJ...517..767G} {517, 767}

\bibitem[\protect\citeauthoryear{{Garc{\'\i}a-Segura}, {L{\'o}pez}  \&
  {Franco}}{{Garc{\'\i}a-Segura} et~al.}{2005}]{Garcia2005}
{Garc{\'\i}a-Segura} G.,  {L{\'o}pez} J.~A.,   {Franco} J.,  2005, \mn@doi
  [\apj] {10.1086/426110}, \href
  {https://ui.adsabs.harvard.edu/abs/2005ApJ...618..919G} {618, 919}

\bibitem[\protect\citeauthoryear{{Garc{\'\i}a-Segura}, {Villaver}, {Manchado},
  {Langer}  \& {Yoon}}{{Garc{\'\i}a-Segura} et~al.}{2016}]{Garcia2016}
{Garc{\'\i}a-Segura} G.,  {Villaver} E.,  {Manchado} A.,  {Langer} N.,   {Yoon}
  S.~C.,  2016, \mn@doi [\apj] {10.3847/0004-637X/823/2/142}, \href
  {https://ui.adsabs.harvard.edu/abs/2016ApJ...823..142G} {823, 142}

\bibitem[\protect\citeauthoryear{{Gies} \& {Bolton}}{{Gies} \&
  {Bolton}}{1986}]{Gies1986}
{Gies} D.~R.,  {Bolton} C.~T.,  1986, \mn@doi [\apjs] {10.1086/191118}, \href
  {https://ui.adsabs.harvard.edu/abs/1986ApJS...61..419G} {61, 419}

\bibitem[\protect\citeauthoryear{{Gull} \& {Sofia}}{{Gull} \&
  {Sofia}}{1979}]{Gull1979}
{Gull} T.~R.,  {Sofia} S.,  1979, \mn@doi [\apj] {10.1086/157137}, \href
  {https://ui.adsabs.harvard.edu/abs/1979ApJ...230..782G} {230, 782}

\bibitem[\protect\citeauthoryear{{Gvaramadze} \& {Gualandris}}{{Gvaramadze} \&
  {Gualandris}}{2011}]{Gvaramadze2011}
{Gvaramadze} V.~V.,  {Gualandris} A.,  2011, \mn@doi [\mnras]
  {10.1111/j.1365-2966.2010.17446.x}, \href
  {https://ui.adsabs.harvard.edu/abs/2011MNRAS.410..304G} {410, 304}

\bibitem[\protect\citeauthoryear{{Gvaramadze}, {R{\"o}ser}, {Scholz}  \&
  {Schilbach}}{{Gvaramadze} et~al.}{2011}]{2011A&A...529A..14G}
{Gvaramadze} V.~V.,  {R{\"o}ser} S.,  {Scholz} R.~D.,   {Schilbach} E.,  2011,
  \mn@doi [\aap] {10.1051/0004-6361/201016256}, \href
  {https://ui.adsabs.harvard.edu/abs/2011A&A...529A..14G} {529, A14}

\bibitem[\protect\citeauthoryear{{Gvaramadze}, {Menten}, {Kniazev}, {Langer},
  {Mackey}, {Kraus}, {Meyer}  \& {Kami{\'n}ski}}{{Gvaramadze}
  et~al.}{2014}]{Gvaramadze2014}
{Gvaramadze} V.~V.,  {Menten} K.~M.,  {Kniazev} A.~Y.,  {Langer} N.,  {Mackey}
  J.,  {Kraus} A.,  {Meyer} D.~M.~A.,   {Kami{\'n}ski} T.,  2014, \mn@doi
  [\mnras] {10.1093/mnras/stt1943}, \href
  {https://ui.adsabs.harvard.edu/abs/2014MNRAS.437..843G} {437, 843}

\bibitem[\protect\citeauthoryear{{Harvey} et~al.,}{{Harvey}
  et~al.}{2020}]{Harvey2020}
{Harvey} E.~J.,  et~al., 2020, \mn@doi [\mnras] {10.1093/mnras/staa2896}, \href
  {https://ui.adsabs.harvard.edu/abs/2020MNRAS.499.2959H} {499, 2959}

\bibitem[\protect\citeauthoryear{{Henden}, {Templeton}, {Terrell}, {Smith},
  {Levine}  \& {Welch}}{{Henden} et~al.}{2016}]{APASS2016}
{Henden} A.~A.,  {Templeton} M.,  {Terrell} D.,  {Smith} T.~C.,  {Levine} S.,
  {Welch} D.,  2016, VizieR Online Data Catalog, \href
  {https://ui.adsabs.harvard.edu/abs/2016yCat.2336....0H} {p. II/336}

\bibitem[\protect\citeauthoryear{{Henney} \& {Arthur}}{{Henney} \&
  {Arthur}}{2019}]{Henney2019a}
{Henney} W.~J.,  {Arthur} S.~J.,  2019, \mn@doi [\mnras]
  {10.1093/mnras/stz1043}, \href
  {https://ui.adsabs.harvard.edu/abs/2019MNRAS.486.3423H} {486, 3423}

\bibitem[\protect\citeauthoryear{{Herbst} et~al.,}{{Herbst}
  et~al.}{2022}]{herbst_ssrv_218_2022}
{Herbst} K.,  et~al., 2022, \mn@doi [\ssr] {10.1007/s11214-022-00894-3}, \href
  {https://ui.adsabs.harvard.edu/abs/2022SSRv..218...29H} {218, 29}

\bibitem[\protect\citeauthoryear{{Hoefner}, {Fleischer}, {Gauger},
  {Feuchtinger}, {Dorfi}, {Winters}  \& {Sedlmayr}}{{Hoefner}
  et~al.}{1996}]{Hoefner1996}
{Hoefner} S.,  {Fleischer} A.~J.,  {Gauger} A.,  {Feuchtinger} M.~U.,  {Dorfi}
  E.~A.,  {Winters} J.~M.,   {Sedlmayr} E.,  1996, \aap, \href
  {https://ui.adsabs.harvard.edu/abs/1996A&A...314..204H} {314, 204}

\bibitem[\protect\citeauthoryear{{H{\o}g} et~al.,}{{H{\o}g}
  et~al.}{2000}]{Tycho2000}
{H{\o}g} E.,  et~al., 2000, \aap, \href
  {https://ui.adsabs.harvard.edu/abs/2000A&A...355L..27H} {355, L27}

\bibitem[\protect\citeauthoryear{{Huthoff} \& {Kaper}}{{Huthoff} \&
  {Kaper}}{2002}]{Huthoff2002}
{Huthoff} F.,  {Kaper} L.,  2002, \mn@doi [\aap] {10.1051/0004-6361:20011793},
  \href {https://ui.adsabs.harvard.edu/abs/2002A&A...383..999H} {383, 999}

\bibitem[\protect\citeauthoryear{{Ishihara} et~al.,}{{Ishihara}
  et~al.}{2010}]{AKARI2010}
{Ishihara} D.,  et~al., 2010, \mn@doi [\aap] {10.1051/0004-6361/200913811},
  \href {https://ui.adsabs.harvard.edu/abs/2010A&A...514A...1I} {514, A1}

\bibitem[\protect\citeauthoryear{{Kaper}, {van Loon}, {Augusteijn},
  {Goudfrooij}, {Patat}, {Waters}  \& {Zijlstra}}{{Kaper}
  et~al.}{1997}]{Kaper1997}
{Kaper} L.,  {van Loon} J.~T.,  {Augusteijn} T.,  {Goudfrooij} P.,  {Patat} F.,
   {Waters} L.~B.~F.~M.,   {Zijlstra} A.~A.,  1997, \mn@doi [\apjl]
  {10.1086/310454}, \href
  {https://ui.adsabs.harvard.edu/abs/1997ApJ...475L..37K} {475, L37}

\bibitem[\protect\citeauthoryear{{Kato}}{{Kato}}{1999}]{HDvar3}
{Kato} T.,  1999, Information Bulletin on Variable Stars, \href
  {https://ui.adsabs.harvard.edu/abs/1999IBVS.4789....1K} {4789, 1}

\bibitem[\protect\citeauthoryear{{Kobulnicky} et~al.,}{{Kobulnicky}
  et~al.}{2016}]{Kobulnicky2016}
{Kobulnicky} H.~A.,  et~al., 2016, \mn@doi [\apjs]
  {10.3847/0067-0049/227/2/18}, \href
  {https://ui.adsabs.harvard.edu/abs/2016ApJS..227...18K} {227, 18}

\bibitem[\protect\citeauthoryear{{Kobulnicky}, {Schurhammer}, {Baldwin},
  {Chick}, {Dixon}, {Lee}  \& {Povich}}{{Kobulnicky}
  et~al.}{2017}]{Kobulnicky2017}
{Kobulnicky} H.~A.,  {Schurhammer} D.~P.,  {Baldwin} D.~J.,  {Chick} W.~T.,
  {Dixon} D.~M.,  {Lee} D.,   {Povich} M.~S.,  2017, \mn@doi [\aj]
  {10.3847/1538-3881/aa90ba}, \href
  {https://ui.adsabs.harvard.edu/abs/2017AJ....154..201K} {154, 201}

\bibitem[\protect\citeauthoryear{{Kukarkin}, {Kholopov}  \&
  {Perova}}{{Kukarkin} et~al.}{1970}]{HDvar2}
{Kukarkin} B.~V.,  {Kholopov} P.~N.,   {Perova} N.~B.,  1970, Information
  Bulletin on Variable Stars, \href
  {https://ui.adsabs.harvard.edu/abs/1970IBVS..480....1K} {480, 1}

\bibitem[\protect\citeauthoryear{{Lamers} \& {Cassinelli}}{{Lamers} \&
  {Cassinelli}}{1999}]{Lamers1999}
{Lamers} H. J.~G.~L.~M.,  {Cassinelli} J.~P.,  1999, {Introduction to Stellar
  Winds}

\bibitem[\protect\citeauthoryear{{Langer}}{{Langer}}{2012}]{Langer2012}
{Langer} N.,  2012, \mn@doi [\araa] {10.1146/annurev-astro-081811-125534},
  \href {https://ui.adsabs.harvard.edu/abs/2012ARA&A..50..107L} {50, 107}

\bibitem[\protect\citeauthoryear{{Lasker} et~al.,}{{Lasker}
  et~al.}{2008}]{Lasker2008}
{Lasker} B.~M.,  et~al., 2008, \mn@doi [\aj] {10.1088/0004-6256/136/2/735},
  \href {https://ui.adsabs.harvard.edu/abs/2008AJ....136..735L} {136, 735}

\bibitem[\protect\citeauthoryear{{L{\'o}pez-Santiago}
  et~al.,}{{L{\'o}pez-Santiago} et~al.}{2012}]{Lopez-Santiago2012}
{L{\'o}pez-Santiago} J.,  et~al., 2012, \mn@doi [\apjl]
  {10.1088/2041-8205/757/1/L6}, \href
  {https://ui.adsabs.harvard.edu/abs/2012ApJ...757L...6L} {757, L6}

\bibitem[\protect\citeauthoryear{{McKee}}{{McKee}}{1995}]{mckee_80_aspc_1995}
{McKee} C.~F.,  1995, in {Ferrara} A.,  {McKee} C.~F.,  {Heiles} C.,
  {Shapiro} P.~R.,  eds,  Astronomical Society of the Pacific Conference Series
  Vol. 80, The Physics of the Interstellar Medium and Intergalactic Medium.
  p.~292

\bibitem[\protect\citeauthoryear{{Meaburn}, {L{\'o}pez}, {Guti{\'e}rrez},
  {Quir{\'o}z}, {Murillo}, {Vald{\'e}z}  \& {Pedrayez}}{{Meaburn}
  et~al.}{2003}]{Meaburn2003}
{Meaburn} J.,  {L{\'o}pez} J.~A.,  {Guti{\'e}rrez} L.,  {Quir{\'o}z} F.,
  {Murillo} J.~M.,  {Vald{\'e}z} J.,   {Pedrayez} M.,  2003, \rmxaa, \href
  {https://ui.adsabs.harvard.edu/abs/2003RMxAA..39..185M} {39, 185}

\bibitem[\protect\citeauthoryear{{Merle}, {Jorissen}, {Van Eck}, {Masseron}  \&
  {Van Winckel}}{{Merle} et~al.}{2016}]{Merle2016}
{Merle} T.,  {Jorissen} A.,  {Van Eck} S.,  {Masseron} T.,   {Van Winckel} H.,
  2016, \mn@doi [\aap] {10.1051/0004-6361/201526944}, \href
  {https://ui.adsabs.harvard.edu/abs/2016A&A...586A.151M} {586, A151}

\bibitem[\protect\citeauthoryear{{Meyer}, {Mackey}, {Langer}, {Gvaramadze},
  {Mignone}, {Izzard}  \& {Kaper}}{{Meyer} et~al.}{2014}]{Meyer2014}
{Meyer} D.~M.~A.,  {Mackey} J.,  {Langer} N.,  {Gvaramadze} V.~V.,  {Mignone}
  A.,  {Izzard} R.~G.,   {Kaper} L.,  2014, \mn@doi [\mnras]
  {10.1093/mnras/stu1629}, \href
  {https://ui.adsabs.harvard.edu/abs/2014MNRAS.444.2754M} {444, 2754}

\bibitem[\protect\citeauthoryear{{Meyer}, {van Marle}, {Kuiper}  \&
  {Kley}}{{Meyer} et~al.}{2016}]{Meyer2016}
{Meyer} D.~M.~A.,  {van Marle} A.~J.,  {Kuiper} R.,   {Kley} W.,  2016, \mn@doi
  [\mnras] {10.1093/mnras/stw651}, \href
  {https://ui.adsabs.harvard.edu/abs/2016MNRAS.459.1146M} {459, 1146}

\bibitem[\protect\citeauthoryear{{Meyer}, {Mignone}, {Kuiper}, {Raga}  \&
  {Kley}}{{Meyer} et~al.}{2017}]{2017MNRAS.464.3229M}
{Meyer} D.~M.~A.,  {Mignone} A.,  {Kuiper} R.,  {Raga} A.~C.,   {Kley} W.,
  2017, \mn@doi [\mnras] {10.1093/mnras/stw2537}, \href
  {https://ui.adsabs.harvard.edu/abs/2017MNRAS.464.3229M} {464, 3229}

\bibitem[\protect\citeauthoryear{{Meyer}, {Oskinova}, {Pohl}  \&
  {Petrov}}{{Meyer} et~al.}{2020}]{meyer_mnras_496_2020}
{Meyer} D.~M.~A.,  {Oskinova} L.~M.,  {Pohl} M.,   {Petrov} M.,  2020, \mn@doi
  [\mnras] {10.1093/mnras/staa1700}, \href
  {https://ui.adsabs.harvard.edu/abs/2020MNRAS.496.3906M} {496, 3906}

\bibitem[\protect\citeauthoryear{{Meyer}, {Mignone}, {Petrov}, {Scherer},
  {Vel{\'a}zquez}  \& {Boumis}}{{Meyer} et~al.}{2021}]{meyer_mnras_506_2021}
{Meyer} D.~M.~A.,  {Mignone} A.,  {Petrov} M.,  {Scherer} K.,  {Vel{\'a}zquez}
  P.~F.,   {Boumis} P.,  2021, \mn@doi [\mnras] {10.1093/mnras/stab2026}, \href
  {https://ui.adsabs.harvard.edu/abs/2021MNRAS.506.5170M} {506, 5170}

\bibitem[\protect\citeauthoryear{{Mohamed}, {Mackey}  \& {Langer}}{{Mohamed}
  et~al.}{2012}]{mohamed_aa_541_2012}
{Mohamed} S.,  {Mackey} J.,   {Langer} N.,  2012, \mn@doi [\aap]
  {10.1051/0004-6361/201118002}, \href
  {http://adsabs.harvard.edu/abs/2012A%26A...541A...1M} {541, A1}

\bibitem[\protect\citeauthoryear{{Neugebauer} et~al.,}{{Neugebauer}
  et~al.}{1984}]{IRAS1984}
{Neugebauer} G.,  et~al., 1984, \mn@doi [\apjl] {10.1086/184209}, \href
  {https://ui.adsabs.harvard.edu/abs/1984ApJ...278L...1N} {278, L1}

\bibitem[\protect\citeauthoryear{{Noriega-Crespo}, {van Buren}  \&
  {Dgani}}{{Noriega-Crespo} et~al.}{1997}]{Noriega-Crespo1997}
{Noriega-Crespo} A.,  {van Buren} D.,   {Dgani} R.,  1997, \mn@doi [\aj]
  {10.1086/118298}, \href
  {https://ui.adsabs.harvard.edu/abs/1997AJ....113..780N} {113, 780}

\bibitem[\protect\citeauthoryear{{Olofsson}, {Maercker}, {Eriksson},
  {Gustafsson}  \& {Sch{\"o}ier}}{{Olofsson} et~al.}{2010}]{Olofsson2010}
{Olofsson} H.,  {Maercker} M.,  {Eriksson} K.,  {Gustafsson} B.,
  {Sch{\"o}ier} F.,  2010, \mn@doi [\aap] {10.1051/0004-6361/200913929}, \href
  {https://ui.adsabs.harvard.edu/abs/2010A&A...515A..27O} {515, A27}

\bibitem[\protect\citeauthoryear{{Peri}, {Benaglia}, {Brookes}, {Stevens}  \&
  {Isequilla}}{{Peri} et~al.}{2012}]{Peri2012}
{Peri} C.~S.,  {Benaglia} P.,  {Brookes} D.~P.,  {Stevens} I.~R.,   {Isequilla}
  N.~L.,  2012, \mn@doi [\aap] {10.1051/0004-6361/201118116}, \href
  {https://ui.adsabs.harvard.edu/abs/2012A&A...538A.108P} {538, A108}

\bibitem[\protect\citeauthoryear{{Peri}, {Benaglia}  \& {Isequilla}}{{Peri}
  et~al.}{2015}]{Peri2015}
{Peri} C.~S.,  {Benaglia} P.,   {Isequilla} N.~L.,  2015, \mn@doi [\aap]
  {10.1051/0004-6361/201424676}, \href
  {https://ui.adsabs.harvard.edu/abs/2015A&A...578A..45P} {578, A45}

\bibitem[\protect\citeauthoryear{{Poveda}, {Ruiz}  \& {Allen}}{{Poveda}
  et~al.}{1967}]{Poveda1967}
{Poveda} A.,  {Ruiz} J.,   {Allen} C.,  1967, Boletin de los Observatorios
  Tonantzintla y Tacubaya, \href
  {https://ui.adsabs.harvard.edu/abs/1967BOTT....4...86P} {4, 86}

\bibitem[\protect\citeauthoryear{{Raga}, {Canto}, {Curiel}  \& {Taylor}}{{Raga}
  et~al.}{1998}]{Raga1998}
{Raga} A.~C.,  {Canto} J.,  {Curiel} S.,   {Taylor} S.,  1998, \mn@doi [\mnras]
  {10.1046/j.1365-8711.1998.01188.x}, \href
  {https://ui.adsabs.harvard.edu/abs/1998MNRAS.295..738R} {295, 738}

\bibitem[\protect\citeauthoryear{{Raskin} et~al.,}{{Raskin}
  et~al.}{2011}]{Raskin2011}
{Raskin} G.,  et~al., 2011, \mn@doi [\aap] {10.1051/0004-6361/201015435}, \href
  {https://ui.adsabs.harvard.edu/abs/2011A&A...526A..69R} {526, A69}

\bibitem[\protect\citeauthoryear{{Schlegel}, {Finkbeiner}  \&
  {Davis}}{{Schlegel} et~al.}{1998}]{NED}
{Schlegel} D.~J.,  {Finkbeiner} D.~P.,   {Davis} M.,  1998, \mn@doi [\apj]
  {10.1086/305772}, \href
  {https://ui.adsabs.harvard.edu/abs/1998ApJ...500..525S} {500, 525}

\bibitem[\protect\citeauthoryear{{Sch{\"o}ier}, {Olofsson}  \&
  {Lundgren}}{{Sch{\"o}ier} et~al.}{2006}]{Schoier2006}
{Sch{\"o}ier} F.~L.,  {Olofsson} H.,   {Lundgren} A.~A.,  2006, \mn@doi [\aap]
  {10.1051/0004-6361:20054795}, \href
  {https://ui.adsabs.harvard.edu/abs/2006A&A...454..247S} {454, 247}

\bibitem[\protect\citeauthoryear{{Shetye} et~al.,}{{Shetye} et~al.}{2018}]{S18}
{Shetye} S.,  et~al., 2018, \mn@doi [\aap] {10.1051/0004-6361/201833298}, \href
  {https://ui.adsabs.harvard.edu/abs/2018A&A...620A.148S} {620, A148}

\bibitem[\protect\citeauthoryear{{Silva Aguirre} et~al.,}{{Silva Aguirre}
  et~al.}{2020}]{Aguirre2020}
{Silva Aguirre} V.,  et~al., 2020, \mn@doi [\aap]
  {10.1051/0004-6361/201935843}, \href
  {https://ui.adsabs.harvard.edu/abs/2020A&A...635A.164S} {635, A164}

\bibitem[\protect\citeauthoryear{{Skrutskie} et~al.,}{{Skrutskie}
  et~al.}{2006}]{2MASS2006}
{Skrutskie} M.~F.,  et~al., 2006, \mn@doi [\aj] {10.1086/498708}, \href
  {https://ui.adsabs.harvard.edu/abs/2006AJ....131.1163S} {131, 1163}

\bibitem[\protect\citeauthoryear{{Smith}}{{Smith}}{2014}]{Smith2014}
{Smith} N.,  2014, \mn@doi [\araa] {10.1146/annurev-astro-081913-040025}, \href
  {https://ui.adsabs.harvard.edu/abs/2014ARA&A..52..487S} {52, 487}

\bibitem[\protect\citeauthoryear{{Smith} \& {Lambert}}{{Smith} \&
  {Lambert}}{1990}]{Smith1990}
{Smith} V.~V.,  {Lambert} D.~L.,  1990, \mn@doi [\apjs] {10.1086/191421}, \href
  {https://ui.adsabs.harvard.edu/abs/1990ApJS...72..387S} {72, 387}

\bibitem[\protect\citeauthoryear{{Spitzer} \& {Varshalovich}}{{Spitzer} \&
  {Varshalovich}}{1980}]{Spitzer1980}
{Spitzer} L.,  {Varshalovich} D.~A.,  1980, Astrofizika, \href
  {https://ui.adsabs.harvard.edu/abs/1980Afz....16..795S} {16, 795}

\bibitem[\protect\citeauthoryear{{Staff}, {De Marco}, {Macdonald}, {Galaviz},
  {Passy}, {Iaconi}  \& {Low}}{{Staff} et~al.}{2016}]{Staff2016}
{Staff} J.~E.,  {De Marco} O.,  {Macdonald} D.,  {Galaviz} P.,  {Passy} J.-C.,
  {Iaconi} R.,   {Low} M.-M.~M.,  2016, \mn@doi [\mnras]
  {10.1093/mnras/stv2548}, \href
  {https://ui.adsabs.harvard.edu/abs/2016MNRAS.455.3511S} {455, 3511}

\bibitem[\protect\citeauthoryear{{Steffen}, {Koning}, {Wenger}, {Morisset}  \&
  {Magnor}}{{Steffen} et~al.}{2011}]{Steffen2011}
{Steffen} W.,  {Koning} N.,  {Wenger} S.,  {Morisset} C.,   {Magnor} M.,  2011,
  \mn@doi [IEEE Transactions on Visualization and Computer Graphics]
  {10.1109/TVCG.2010.62}, \href
  {https://ui.adsabs.harvard.edu/abs/2011ITVCG..17..454S} {17, 454}

\bibitem[\protect\citeauthoryear{{Stevens}, {Blondin}  \& {Pollock}}{{Stevens}
  et~al.}{1992}]{1992ApJ...386..265S}
{Stevens} I.~R.,  {Blondin} J.~M.,   {Pollock} A.~M.~T.,  1992, \mn@doi [\apj]
  {10.1086/171013}, \href
  {https://ui.adsabs.harvard.edu/abs/1992ApJ...386..265S} {386, 265}

\bibitem[\protect\citeauthoryear{{Toal{\'a}}, {Guerrero}, {Ramos-Larios}  \&
  {Guzm{\'a}n}}{{Toal{\'a}} et~al.}{2015}]{Toala2015}
{Toal{\'a}} J.~A.,  {Guerrero} M.~A.,  {Ramos-Larios} G.,   {Guzm{\'a}n} V.,
  2015, \mn@doi [\aap] {10.1051/0004-6361/201525706}, \href
  {https://ui.adsabs.harvard.edu/abs/2015A&A...578A..66T} {578, A66}

\bibitem[\protect\citeauthoryear{{Villaver} \& {Stanghellini}}{{Villaver} \&
  {Stanghellini}}{2005}]{villaver_apj_632_2005}
{Villaver} E.,  {Stanghellini} L.,  2005, \mn@doi [\apj] {10.1086/433183},
  \href {https://ui.adsabs.harvard.edu/abs/2005ApJ...632..854V} {632, 854}

\bibitem[\protect\citeauthoryear{{Villaver}, {Manchado}  \&
  {Garc{\'\i}a-Segura}}{{Villaver} et~al.}{2012}]{villalver_apj_748_2012}
{Villaver} E.,  {Manchado} A.,   {Garc{\'\i}a-Segura} G.,  2012, \mn@doi [\apj]
  {10.1088/0004-637X/748/2/94}, \href
  {https://ui.adsabs.harvard.edu/abs/2012ApJ...748...94V} {748, 94}

\bibitem[\protect\citeauthoryear{{Vishniac}}{{Vishniac}}{1994}]{vishniac_apj_428_1994}
{Vishniac} E.~T.,  1994, \mn@doi [\apj] {10.1086/174231}, \href
  {http://adsabs.harvard.edu/abs/1994ApJ...428..186V} {428, 186}

\bibitem[\protect\citeauthoryear{{Wareing}, {Zijlstra}, {O'Brien}  \&
  {Seibert}}{{Wareing} et~al.}{2007}]{Wareing2007}
{Wareing} C.~J.,  {Zijlstra} A.~A.,  {O'Brien} T.~J.,   {Seibert} M.,  2007,
  \mn@doi [\apjl] {10.1086/524407}, \href
  {https://ui.adsabs.harvard.edu/abs/2007ApJ...670L.125W} {670, L125}

\bibitem[\protect\citeauthoryear{{Weaver}, {McCray}, {Castor}, {Shapiro}  \&
  {Moore}}{{Weaver} et~al.}{1977}]{Weaver1977}
{Weaver} R.,  {McCray} R.,  {Castor} J.,  {Shapiro} P.,   {Moore} R.,  1977,
  \mn@doi [\apj] {10.1086/155692}, \href
  {https://ui.adsabs.harvard.edu/abs/1977ApJ...218..377W} {218, 377}

\bibitem[\protect\citeauthoryear{{Weigelt}, {Balega}, {Bl{\"o}cker}, {Hofmann},
  {Men'shchikov}  \& {Winters}}{{Weigelt} et~al.}{2002}]{Weigelt2002}
{Weigelt} G.,  {Balega} Y.~Y.,  {Bl{\"o}cker} T.,  {Hofmann} K.~H.,
  {Men'shchikov} A.~B.,   {Winters} J.~M.,  2002, \mn@doi [\aap]
  {10.1051/0004-6361:20020915}, \href
  {https://ui.adsabs.harvard.edu/abs/2002A&A...392..131W} {392, 131}

\bibitem[\protect\citeauthoryear{{Wenger} et~al.,}{{Wenger} et~al.}{2000}]{CDS}
{Wenger} M.,  et~al., 2000, \mn@doi [\aaps] {10.1051/aas:2000332}, \href
  {https://ui.adsabs.harvard.edu/abs/2000A&AS..143....9W} {143, 9}

\bibitem[\protect\citeauthoryear{{Wilkin}}{{Wilkin}}{1996}]{wilkin_459_apj_1996}
{Wilkin} F.~P.,  1996, \mn@doi [\apjl] {10.1086/309939}, \href
  {http://adsabs.harvard.edu/abs/1996ApJ...459L..31W} {459, L31}

\bibitem[\protect\citeauthoryear{{Wilkin} \& {Stahler}}{{Wilkin} \&
  {Stahler}}{1996}]{Wilkin1996}
{Wilkin} F.~P.,  {Stahler} S.~W.,  1996, in American Astronomical Society
  Meeting Abstracts. p. 53.05

\bibitem[\protect\citeauthoryear{{Wright} et~al.,}{{Wright}
  et~al.}{2010}]{Wright2010}
{Wright} E.~L.,  et~al., 2010, \mn@doi [\aj] {10.1088/0004-6256/140/6/1868},
  \href {https://ui.adsabs.harvard.edu/abs/2010AJ....140.1868W} {140, 1868}

\bibitem[\protect\citeauthoryear{{Zwicky}}{{Zwicky}}{1957}]{Zwicky1957}
{Zwicky} F.,  1957, {Morphological astronomy}

\bibitem[\protect\citeauthoryear{{van Buren}, {Noriega-Crespo}  \&
  {Dgani}}{{van Buren} et~al.}{1995}]{vanBuren1995}
{van Buren} D.,  {Noriega-Crespo} A.,   {Dgani} R.,  1995, \mn@doi [\aj]
  {10.1086/117739}, \href
  {https://ui.adsabs.harvard.edu/abs/1995AJ....110.2914V} {110, 2914}

\bibitem[\protect\citeauthoryear{{van Marle}, {Meliani}, {Keppens}  \&
  {Decin}}{{van Marle} et~al.}{2011}]{vanmarle_apj_734_2011}
{van Marle} A.~J.,  {Meliani} Z.,  {Keppens} R.,   {Decin} L.,  2011, \mn@doi
  [\apjl] {10.1088/2041-8205/734/2/L26}, \href
  {https://ui.adsabs.harvard.edu/abs/2011ApJ...734L..26V} {734, L26}

\bibitem[\protect\citeauthoryear{{van Marle}, {Meliani}  \& {Marcowith}}{{van
  Marle} et~al.}{2012}]{vanMarle2012}
{van Marle} A.~J.,  {Meliani} Z.,   {Marcowith} A.,  2012, \mn@doi [\aap]
  {10.1051/0004-6361/201219180}, \href
  {https://ui.adsabs.harvard.edu/abs/2012A&A...541L...8V} {541, L8}

\bibitem[\protect\citeauthoryear{{van Marle}, {Meliani}  \& {Marcowith}}{{van
  Marle} et~al.}{2015}]{vanMarle2015}
{van Marle} A.~J.,  {Meliani} Z.,   {Marcowith} A.,  2015, \mn@doi [\aap]
  {10.1051/0004-6361/201425230}, \href
  {https://ui.adsabs.harvard.edu/abs/2015A&A...584A..49V} {584, A49}

\makeatother
\end{thebibliography}

\end{document}